\documentclass[aps,pra,reprint,superscriptaddress,amssymb,floatfix]{revtex4-2}

\usepackage{graphicx}
\usepackage{subfigure}
\usepackage{amsthm}
\usepackage{amsmath}
\usepackage{bbm}

\usepackage{algorithm}
\usepackage{algorithmic}

\RequirePackage[colorlinks=true,linkcolor=blue,urlcolor=blue,citecolor=blue]{hyperref}

\newcommand{\nc}{\newcommand}
\nc{\n}{\operatorname}
\newcommand{\QCAEmodel}{QCAE}
\newcommand{\VQAname}{{\em varQCAE}}
\newcommand{\MaxMixedState}{\omega}
\newcommand{\MaxEntangledState}{\phi^+}

\nc{\Cal}[1]{\mathcal{#1}} \nc{\fr}[1]{\mathfrak{#1}}
\nc{\Ac}{\Cal{A}} \nc{\Bc}{\Cal{B}} \nc{\Cc}{\Cal{C}} \nc{\Dc}{\Cal{D}} \nc{\Ec}{\Cal{E}}
\nc{\Fc}{\Cal{F}} \nc{\Gc}{\Cal{G}} \nc{\Hc}{\Cal{H}} \nc{\Ic}{\Cal{I}} \nc{\Jc}{\Cal{J}}
\nc{\Kc}{\Cal{K}} \nc{\Lc}{\Cal{L}} \nc{\Mc}{\Cal{M}} \nc{\Nc}{\Cal{N}} \nc{\Oc}{\Cal{O}}
\nc{\Pc}{\Cal{P}} \nc{\Qc}{\Cal{Q}} \nc{\Rc}{\Cal{R}} \nc{\Sc}{\Cal{S}} \nc{\Tc}{\Cal{T}}
\nc{\Uc}{\Cal{U}} \nc{\Vc}{\Cal{V}} \nc{\Wc}{\Cal{W}} \nc{\Xc}{\Cal{X}} \nc{\Yc}{\Cal{Y}}
\nc{\Zc}{\Cal{Z}} 
\nc{\Tr}{\n{Tr}} \nc{\tr}{\n{tr}}
\newcommand{\id}{\textit{id}} \newcommand{\1}{\mathbbm{1}}
\newcommand{\ot}{\otimes}  
\nc{\tht}{\theta} \nc{\om}{\omega}

\nc{\brak}[1]{\langle{#1}\rangle}			\nc{\ketb}[2]{|{#1}\rangle\!\langle{#2}|} 
\nc{\bra}[1]{\langle{#1}|} \nc{\ket}[1]{|{#1}\rangle}
\nc{\too}{\!\!\to\!\!} 
\nc{\im}{\n{im}}

\nc{\be}[2]{\begin{#1}#2\end{#1}}
\newtheorem{thm}{Theorem}					\nc{\Thm}[1]{\be{thm}{#1}}
\newtheorem{prop}[thm]{Proposition}			\nc{\Prop}[1]{\be{prop}{#1}}
			\nc{\Coro}[1]{\be{coro}{#1}}
\newtheorem{lem}[thm]{Lemma}				\nc{\Lem}[1]{\be{lem}{#1}}

\begin{document}


\title{Quantum Circuit AutoEncoder}

\author{Jun Wu}%
\email{jun_wu@mail.ustc.edu.cn}
\affiliation{University of Science and Technology of China, Hefei 230027, China }
\author{Hao Fu}
\affiliation{University of Science and Technology of China, Hefei 230027, China }
\author{Mingzheng Zhu}
\affiliation{University of Science and Technology of China, Hefei 230027, China }
\author{Haiyue Zhang}
\affiliation{University of Illinois Urbana-Champaign, Urbana-Champaign, IL, 61801}
\author{Wei Xie}%
\email{xxieww@ustc.edu.cn }
\thanks{Corresponding author}
\affiliation{University of Science and Technology of China, Hefei 230027, China }
\author{Xiang-Yang Li}%
\email{xiangyangli@ustc.edu.cn}
\thanks{Corresponding author}
\affiliation{University of Science and Technology of China, Hefei 230027, China }
\affiliation{Hefei National Laboratory, University of Science and Technology of China, Hefei 230088, China}


\date{\today}

\begin{abstract}
Quantum autoencoder is a quantum neural network model for compressing information stored in quantum states.
However, one needs to process information stored in quantum circuits for many tasks in the emerging quantum information technology.
In this work, generalizing the ideas of classical and quantum autoencoder, we introduce the model of Quantum Circuit AutoEncoder (\QCAEmodel) to compress and encode information within quantum circuits.
We provide a comprehensive protocol for \QCAEmodel~ and design a variational quantum algorithm, \VQAname, for its implementation.
We theoretically analyze this model by deriving conditions for lossless compression and establishing both upper and lower bounds on its recovery fidelity.
Finally, we apply \VQAname~ to three practical tasks and numerical results show that it can effectively (1) compress the information within quantum circuits, (2) detect anomalies in quantum circuits, and (3) mitigate the depolarizing noise in quantum devices.
This suggests that our algorithm is potentially applicable to other information processing tasks for quantum circuits.
\end{abstract}


\maketitle
\section{Introduction}
\begin{figure*}[!htbp]
    \centering
    \subfigure[Autoencoder]{
    \label{fig:AE-frame}
    \centering
    \includegraphics[width=0.31\textwidth]{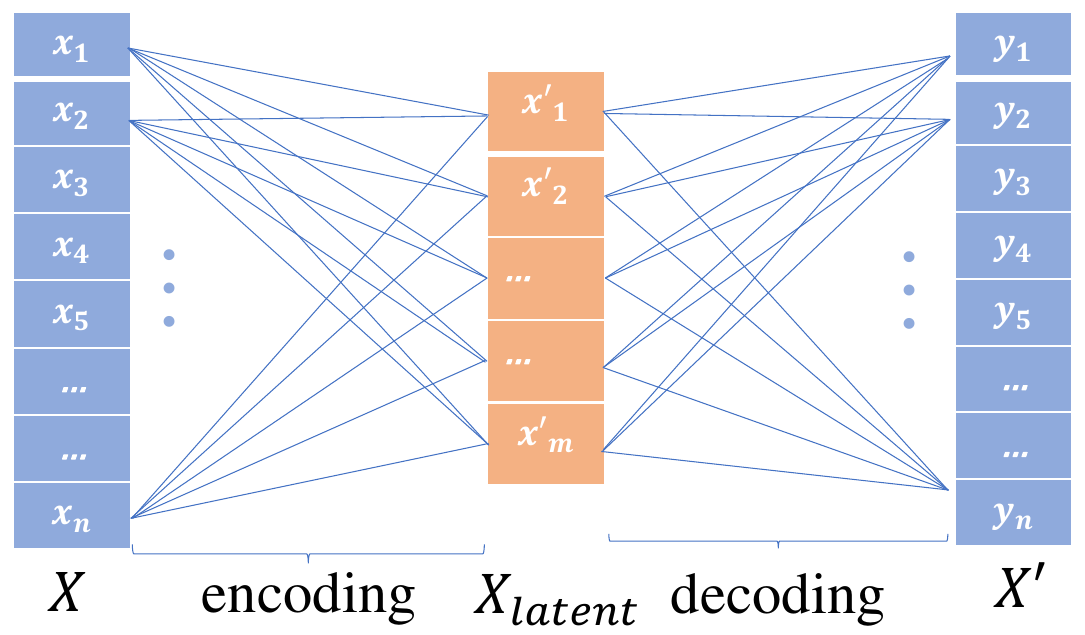}
    }
    \subfigure[Quantum autoencoder]{
    \label{fig:QAE-frame}
    \centering
    \includegraphics[width=0.31\textwidth]{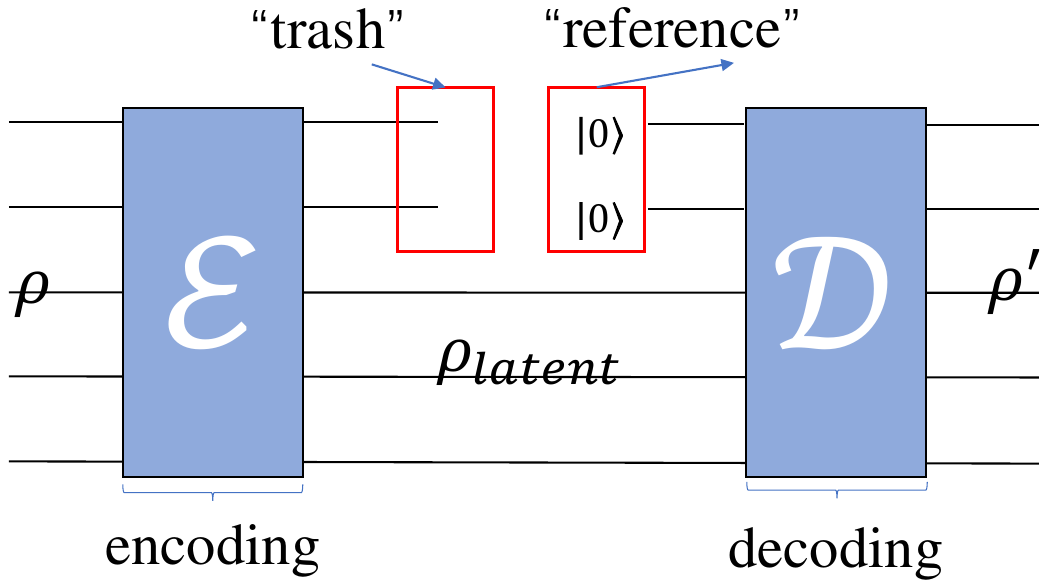}
    }
    \subfigure[Quantum circuit autoencoder]{
    \label{fig:QCAE-frame}
    \centering
    \includegraphics[width=0.31\textwidth]{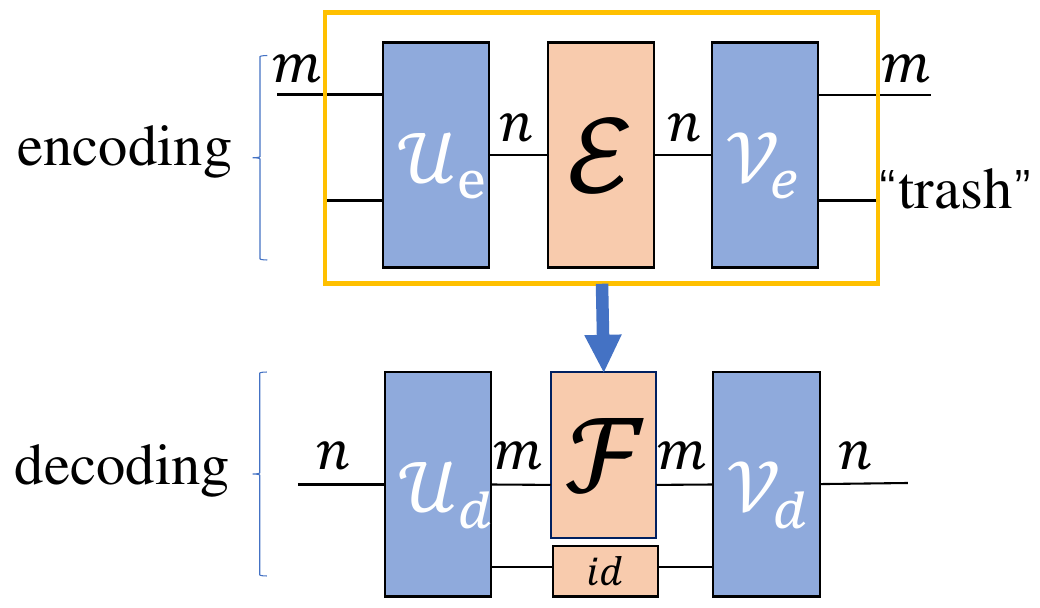}
    }
    \caption{The diagrams of three different autoencoders. (a) The standard autoencoder encodes an $n$-dimensional input data $X$ into a lower dimensional representation $X_{latent}$ of dimensionality $m$, which is then decoded to reconstruct the original $n$-dimensional data as $X'$. (b) The quantum autoencoder takes an input $n$-qubits state $\rho$, transforms it into a lower-dimensional($m$-qubits) state $\rho_{latent}$, and then decodes this state to an $n$-dimensional state $\rho'$. (c) The quantum circuit autoencoder encodes a $D$-dimensional quantum circuit into a $d$-dimensional circuit and reconstructs the original $D$-dimensional quantum circuit through the encoding process.}
    \label{fig:(Q)AE-frame}
\end{figure*}

Autoencoder is a prevalent artificial neural network approach for compressing and encoding information \cite{liou2008modeling}. In Fig.~\ref{fig:AE-frame}, a typical autoencoder framework is depicted, showcasing the primary concept of information compression through a bottleneck while preserving data reconstruction fidelity. Notably, a quantum autoencoder (QAE) has been proposed \cite{romero2017quantum}, extensively explored in quantum machine learning and related domains \cite{wan2017quantum, verdon2018universal, bondarenko2020quantum, huang2020realization, du2021exploring, cerezo2021cost}.

The QAE methodology involves information compression by discarding the ``trash" system during the encoding step, followed by state reconstruction aided by a ``reference" state. Fig.~\ref{fig:QAE-frame} illustrates the typical diagram of a quantum autoencoder, showcasing its process for efficient quantum information compression and reconstruction. However, QAE has a limited fidelity bound when dealing with a large number of input states \cite{cao2021noise}. Furthermore, the quantum information processing for quantum circuits other than quantum states is also a common practice \cite{giovannetti2008quantum, bharti2022noisy}. Classical information is often converted to quantum information in certain quantum-machine-learning tasks through a parameterized encoding circuit. For instance, Ref.~\cite{grant2018hierarchical} employs 4-qubit circuits to transform the Iris dataset (a public image dataset)\cite{misc_iris_53} into quantum states, with information stored both in quantum states and circuits. QAE cannot be directly applied in compressing the information stored within the quantum circuit.

Considering the issues above, there is a need for an elaborate study on quantum circuit autoencoder. The quantum circuit autoencoder can also act as a generalization of QAE. For example, it can subsume QAE in some cases, such as the purified quantum query access model. 

Ref.~\cite{chiribella2015universal} proposed a gate compression model that uses two unitary operators to reduce the input gate's dimension and another two unitary operators to reconstruct the original gate. The authors also provided a method to achieve exponential reduction in dimension. Ref.~\cite{zhu2023quantum} applied quantum autoencoder on quantum cloud computing, proposed a quantum gate autoencoder for reducing the communication qubit resources. These two models can be considered as a prototype of the quantum circuit autoencoder. However, they only consider quantum circuits consisting of single-qubit gates in the form of IID and a family of parameterized quantum circuits, whereas general quantum circuits may consist of multiple qubits and not just single-qubit gates. 

In this paper, we propose a quantum circuit autoencoder model(\QCAEmodel) as depicted in Fig.~\ref{fig:QCAE-frame}. For a quantum channel $\Ec$ acting on $n$-qubits, we construct encoders $\Uc_e$ and $\Vc_e$ to obtain $\Fc = tr_{trash}[\Vc_e \circ \Ec \circ \Uc_e]$ acting on $m$-qubit system ($m < n$), where $\Uc_e$, $\Vc_e$ and partial trace operation consist a supermap \cite{chiribella2008transforming} that maps an $n$-qubits channel to an $m$-qubits channel. The goal is to maximize the reconstruction fidelity between $\tilde{\Ec} = \Vc_d \circ [\Fc \otimes \id] \circ \Uc_d$ and $\Ec$, where $\id$ is the identity channel.  

To implement the \QCAEmodel~on NISQ devices, we design a variational quantum algorithm (VQA) \cite{cerezo2021variational}, referred to as \VQAname. By setting the encoders and decoders as the parameterized quantum circuits(PQCs) \cite{benedetti2019parameterized}, we use the classical optimizer to find optimal parameters for the quantum circuit autoencoder, obtaining executable sequences of local gates suitable for NISQ devices. A VQA consists of PQCs, loss function, and optimizer, and an inevitable issue is the Barren Plateau (BP) \cite{mcclean2018barren}. We use the hardware efficient ansatz\cite{kandala2017hardware} as the PQCs in \VQAname. We propose a perfect compression condition that can help design the loss function to decrease the computation cost. A local cost function, inspired by Ref.~\cite{cerezo2021cost}, is also designed to reduce the impact of BP. Furthermore, we analyze the fidelity bound of \VQAname, including an upper bound for general channels and a lower bound for a special case.

Conventional autoencoders have diverse applications, such as dimension reduction \cite{wang2016auto}, anomaly detection \cite{chalapathy2019deep}, and denoising \cite{gondara2016medical}. Our work employs \VQAname~for quantum circuit tasks, including information compression, anomaly detection, and denoising on quantum circuits. We evaluate the performance of \VQAname~on IBM qiskit \cite{Qiskit} and Mindquantum \cite{mq_2021}. In our experiments, the \VQAname~can compress the information within parameterized quantum circuits with a reconstruction error of approximately 0.05. Moreover, the distribution of anomalous scores of ``normal" and ``abnormal" quantum circuits datasets are significantly different, in which we use two different ways to generate circuits in these two datasets. As for denoising, \VQAname~can reduce the impact of depolarizing error on circuits. In summary, these results indicate that \VQAname~has the potential to be applied to these applications.

\section{Preliminary\label{prelimi}}
A quantum system $A$ corresponds to a Hilbert space $\Hc_A$. The {\em quantum state} of system $A$ is described by a density operator on $\Hc_A$, which is a positive semidefinite operator with trace one.
A quantum state $\rho$ is called {\em pure} if it has rank one and called {\em mixed} otherwise.

In this work, we denote the {\em maximally mixed state} as $\MaxMixedState = \1/d$ and the {\em maximally entangled state} as $\MaxEntangledState = (1/d) \sum_{i,j=0}^{d-1} 
|i\rangle \langle j| \otimes |i\rangle \langle j|$ for a $d^2$-dimensional system. The {\em fidelity} between two quantum states $\rho$ and $\sigma$ is defined as
\begin{equation}
    \label{Eq: fidelity_function}
    F(\rho,\sigma) := \|\sqrt\rho\sqrt\sigma\|_1^2
    =\Big(tr\sqrt{\sqrt\rho\sigma\sqrt\rho}\Big)^2,
\end{equation}
with a special case $F(\rho,|\psi\rangle \langle\psi|)=\langle\psi|\rho|\psi\rangle$.

A {\em quantum operation} (or {\em quantum channel}) $\Ec_{A \to B}$ with input system $A$ and output system $B$ is a completely positive, trace-preserving linear map from the linear operators on $\Hc_A$ to the linear operators on $\Hc_B$. We use $\id$ to denote the identity quantum channel, which means $\id (\rho) = \rho$ for any state $\rho$. The {\em mixed-unitary quantum channel} is defined as the convex combination of unitary operations. For a series of quantum circuits $U_1, U_2, \dots$, we can utilize a controlled circuit to implement a mixed-unitary channel in practice \cite{wei2018efficient}.

In this work, subscripts indicate the input and output systems, and we omit the identity operator $\1$ when it does not introduce ambiguity. 
For instance, $X_AY_B\equiv Y_BX_A\equiv X_A\otimes Y_B$ denotes applying $X_A\otimes Y_B$ to the composite system $AB$.
We write $X_{AB}Y_{BC}\equiv (X_{AB}\otimes \1_C)(\1_A\otimes Y_{BC})$ and $\Ec_{B\to C}(X_{AB})\equiv (\id_A\otimes\Ec_{B\to C})X_{AB}$.
We also write the partial trace of a multipartite operator by directly omitting the subscript the partial trace takes on, for example, $X_B:=tr_A(X_{AB})$.

A quantum channel can be represented by a {\em Choi state} \cite{jamiolkowski1972linear,choi1975completely}. The Choi state of a quantum operation $\Ec_{A\to B}$ is defined as
\begin{equation}
\label{Eq: choi}
\begin{aligned}
     J^\Ec & = (\id_{\tilde A\to A}\ot\Ec_{A\to B})\MaxEntangledState_{\tilde A A} \\
    & =\frac{1}{d_A}\sum_{i,j=0}^{d_A-1}\ketb{i}{j}\ot\Ec(\ketb{i}{j}),
\end{aligned}
\end{equation}
where $\tilde A,A$ are isomorphic systems, and $\{ \ket{i}\}$ is an orthogonal basis of the input space $\Hc_A$.

The output of the channel $\Ec_{A\to B}$ with input $\rho_A$ can be recovered by
\begin{equation}
\label{evolution_of_choi}
    \Ec_{A\to B}(\rho_A)=d_A tr_A(J^\Ec(\rho_A^\intercal \ot\1_B)).
\end{equation}

Let $\Pi_{A_1A_2 \to B_1B_2}$ be a quantum channel with Choi state $J^\Pi$. Then its reduced channel $\Nc_{A_1 \to B_1}$ can be defined as the channel that has Choi state $tr_{A_2B_2}(J^\Pi)$.

In this work, the similarity between two quantum channels is characterized by the fidelity of their respective Choi states. 
To be specific, we define the fidelity of two quantum channels $\Ec_1$ and $\Ec_2$,
\begin{equation}
\label{eq:channel_fidelity}
    F(\Ec_1, \Ec_2) := F(J^{\Ec_1}, J^{\Ec_2}),
\end{equation}
where the right $F(\cdot)$ is the fidelity function defined as in Eq.~(\ref{Eq: fidelity_function}).

\section{Method}\label{Sec:method}

\subsection{Sketch of our method}
\label{sec: method_sketch}
We present the diagram of our \QCAEmodel~model. The goal is to find encoders and decoders to encode $\Ec$ through a bottleneck and decode it to original circuits as faithfully as possible. We design the \VQAname, a variational quantum algorithm, to implement \QCAEmodel. Our algorithm uses the parameterized quantum circuits controlled by a set of parameters to represent the encoders and decoders. Therefore, \VQAname~aims to find the optimal control parameters to maximize the similarity between original and reconstructed quantum channels. 

The \QCAEmodel, as shown in Fig.~\ref{fig:QCAE-frame},
consists of two separate processes: encoding and decoding. During the encoding process, the training dataset $\{U_i\}_{i=1}^{N_{train}}$ is encoded as a mixed-unitary quantum channel $\Ec$ on $n$-qubits system. For an arbitrary state $\rho$, the mixed-unitary quantum channel $\Ec$ can be written as
\begin{equation}
\label{eq: mixed-channel}
    \Ec(\rho) = \sum_{i=1}^{N_{train}} p_i U_i \rho U_i^\dagger.
\end{equation}
Then, the encoders $\Uc_e(\tht)$ and $\Vc_e(\tht)$ act on the channel $\Ec$ and obtain the reduced channel $\Fc$ by partially tracing the last $(n-m)$ qubits(i.e., ``trash" systems). As a result, the encoders and partial trace together form an operator supermaps a $2^n$-dimensional channel $\Ec$ to a $2^m$-dimensional channel $\Fc$. In the decoding process, the decoders $\Uc_d(\tht)$ and $\Vc_d(\tht)$ are applied to the channel $\Fc\ot\id$ to yield a new quantum channel $\Tilde{\Ec}$. Finally, the similarity between $\Ec$ and $\Tilde{\Ec}$ is feed to the classical optimizer to update parameters $\tht$, and repeat the same procedure until the loss function convergence or satisfy other termination conditions. 

\begin{figure*}[!htbp]
    \centering
    \includegraphics[width=0.9\textwidth]{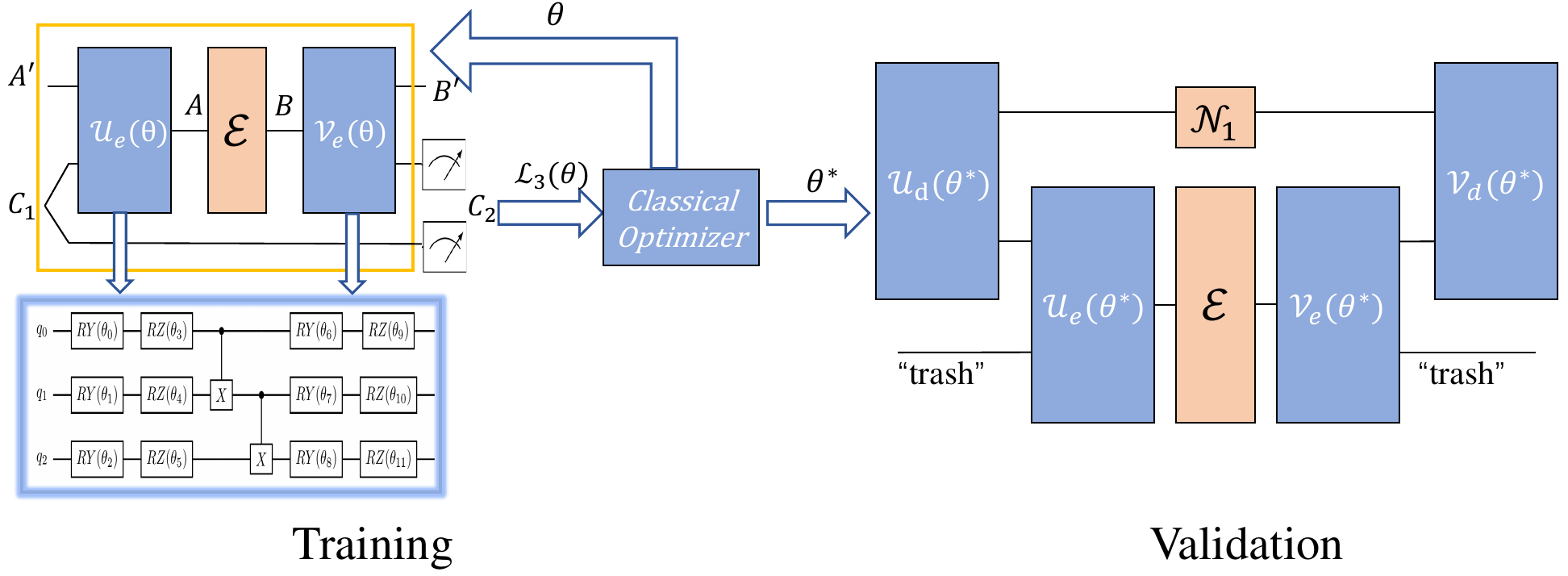}
    \caption{The diagram of the \VQAname. In the training process of \VQAname, we only consider the encoding process in the whole \QCAEmodel framework in Fig.~\ref{fig:QCAE-frame}, as we set the decoders as the dagger of the encoders. We apply the quantum channel $\Pi := \Vc_e \circ \Ec \circ \Uc_e$ on the initial state $\MaxMixedState \ot \MaxEntangledState$. The resulting state is on the composite system $B'C_2$, and we compare the state fidelity between two states on the subsystems $C_1$ and $C_2$. The loss function estimates the state fidelity discussed in Sec.~\ref{subsection: lossfunction}. The near-optimal parameter $\tht^\star$ is obtained after the training and sent to the decoding process of \QCAEmodel. After decoding, the reconstructed channel $\Tilde{\Ec}$ is obtained. The validation process compares the fidelity between $\Ec$ and $\Tilde{\Ec}$. (Note: the systems $C_1$ contains two isomorphic registers, can be written as $C_1 C_1'$, and by the same token $C_2$ can be written as $C_2 C_2'$, The quantum channel $\Pi$ is only applied on the systems $A'C_1$, and we omit $C_1'$ and $C_2'$ for convenient.)}
    \label{fig:QCAE-digram}
\end{figure*}

In the \VQAname, the decoders in the decoding process is set to be the conjugate transpose of the encoders, i.e., $\Uc_d = \Uc_e^\dagger$ and $\Vc_d = \Vc_e^\dagger$. Therefore, we could only consider the encoding process and omit the subscript in encoders and decoders for convenience. As shown in Fig.~\ref{fig:QCAE-frame}, the channel $\Pi := \Vc(\tht) \circ \Ec \circ \Uc(\tht)$ is applied on the product state $\MaxMixedState \ot \MaxEntangledState$, where $\MaxMixedState$ is the maximally mixed state and $\MaxEntangledState$ is the maximally entangled state. The reason for using this initial state is under the consideration of the construction of the loss function, which will be explained in Sec.~\ref{subsection: lossfunction} and Prop.~\ref{perfect_compress_recovery_theorem}. The resulting state is on the composite system $B' C_2$ where the subsystem $C_2$ is defined as the ``trash" system. Considering utilizing the decoding scheme as the validation, the decoders are applied on $\Fc\ot\id$, which is under the assumption that the channel $\Pi_{C_1 C_2}$ is the identity channel. As a sequence, the goal of \VQAname~ is to maximize the fidelity of $\MaxEntangledState_{C_1}$ and $\MaxEntangledState_{C_2}$. Moreover, we propose a condition to compress a quantum channel faithfully; see Prop.~\ref{perfect_compress_recovery_theorem} for more. Then, we use the classical optimizer to optimize the control parameters. 
The hardware-efficient ansatz \cite{kandala2017hardware} is used as ansatzes in our variational algorithm.
Check Alg.~\ref{alg: VQCC_frame} and Fig.~\ref{fig:QCAE-digram} for more details of \VQAname.
(Note: the systems $C_1$ contains two isomorphic registers, can be written as $C_1 C_1'$, and by the same token $C_2$ can be written as $C_2 C_2'$, The quantum channel $\Pi$ is only applied on the systems $A'C_1$, and we omit $C_1'$ and $C_2'$ for convenient.)

\begin{algorithm}[H]
    \label{alg: VQCC_frame}
    \textbf{Input}: Training data $\Dc_{train} = \{U_1, \cdots, U_n \}$, circuit ansatzes $\mathcal{U}(\theta)$ and $\mathcal{V}(\theta)$ and the number of iterations $ITR$;\\
    \begin{algorithmic}[1]
        \STATE \textbf{Training Process:}
        \STATE Set epoch $t=0$ and initialize $\tht_0$ randomly;
        \STATE Let encoders $\Uc_e(\tht) \gets \Uc(\tht)$ and $\Vc_e(\tht) \gets \Vc(\tht)$;
        \WHILE{not converged or $t \leqslant ITR$ }
        \STATE $t \gets t+1$;
        \STATE Initialize loss $\Lc(\tht_{t}) = 0$;
        \FOR {each $U_i$ in $\Dc_{train}$}
        \STATE $\Pi \gets \Vc(\tht_{t}) \circ U_i \circ \Uc(\tht_{t})$;
        \STATE $\psi_{B' C_2} \gets \Pi (\MaxMixedState_{A'} \ot \MaxEntangledState_{C_1})$;
        \STATE $\Lc_3^i(\tht_{it}) \gets \tr[O \tr_{B'}[\psi_{B' C_2}]]$, where $O$ is the observable in $\Lc_3$ that is shown in Eq.~(\ref{eq: local_observable});
        \STATE $\Lc(\tht_{it}) \gets \Lc(\tht_{it}) + \frac{1}{n} [ 1-\Lc_3^i(\tht_{it})]$;
        \ENDFOR
        \STATE Update parameters $\tht_{it+1}$ of $\Lc(\cdot)$ using classical optimizer; 
        \ENDWHILE
        \STATE \textbf{Output} the near-optimal parameters $\tht^\star$;
        
        \STATE \textbf{Validation Process:}
        \STATE Initialize the validation $L_{val} = 0$;
        \STATE Let decoders $\Uc_d(\tht) \gets \Uc(\tht)^\dagger$ and $\Vc_d(\tht) \gets \Vc(\tht)^\dagger$;
        \FOR {each $U_i$ in $\Dc_{train}$}
        \STATE Apply encoders to obtain $\Fc_i \gets tr_{C_1 C_2} [\Vc_e(\tht^\star) \circ U_i \circ \Uc_e(\tht^\star)]$;
        \STATE $\Tilde{U_i} \gets \Vc(\tht^\star)^\dagger \circ (\Fc_i \ot \id) \circ $ $\Uc(\tht^\star)^\dagger$;
        \STATE Calculate the reconstruction fidelity $F(J^{U_i}, J^{\Tilde{U_i}})$;
        \STATE $L_{val} \gets L_{val} + \frac{1}{n}(1-F(J^{U_i}, J^{\Tilde{U_i}}))$;
        \ENDFOR
    \STATE \textbf{Output} The validation value $L_{val}$;
    \end{algorithmic}
    \caption{Main Algorithm: \VQAname}
\end{algorithm}

\subsection{Loss function}
\label{subsection: lossfunction}
In this section, we define the loss function of the train process in \VQAname. From the description in Sec.~\ref{sec: method_sketch}, we know that the goal of \QCAEmodel~is to maximize the reconstruction fidelity between the original channel $\Ec$ and the reconstruction channel $\Tilde{\Ec}$. We first consider the reconstruction fidelity as the loss function. Given the data set $\Dc := \{U_m \}_{m=1}^N$ and parameterized quantum circuits $U(\tht)$ and $V(\tht)$, the loss function is designed to be the mean square error:
\begin{equation}
    \label{Eq:mse_loss}
    \begin{aligned}
    \Lc_1 (\Dc, \tht) := 1-\frac{1}{N} \sum_{m=1}^N \left[ F(U_m, \Tilde{U}_m) \right]^2,
    \end{aligned}
\end{equation}
where $F(\cdot)$ is the fidelity function as defined in Eq.~(\ref{eq:channel_fidelity}), $\Tilde{U}_m = \Vc^\dagger \circ (\Fc_m \ot \id) \circ \Uc^\dagger$ and $\Fc_m = \tr_{C_1 C_2} [\Vc \circ U_m \circ \Uc]$, and the partial trace is on the subsystem $C_1 C_2$ as shown in Fig.~\ref{fig:QCAE-digram}.

$\Lc_1$ is the mean square error of the reconstruction fidelity for each input quantum circuit $U_m$. Additionally, it is consistent with the definition of the mixed quantum channel in Eq.~(\ref{eq: mixed-channel}). More specifically, in lines 4-12 in the Alg.~\ref{alg: VQCC_frame}, we use an equivalent description to the Eq.~(\ref{eq: mixed-channel}). 
\begin{equation}
\label{eq: mixed_channel_in_alg}
\begin{aligned}
    \left[\Vc \circ \Ec \circ \Uc \right](\rho) 
    & = \Vc \circ \Ec (\Uc (\rho))\\
    & = \Vc (\sum_i p_i U_i (\Uc (\rho)) U_i^\dagger) \\
    & = \sum_i p_i \Vc (U_i \Uc (\rho) U_i^\dagger).
\end{aligned}
\end{equation} 
As a result, $\Lc_1$ comprehensively evaluates the training process. However, for several reasons, $\Lc_1$ falls short as a proper loss function. Firstly, the loss function in Eq.~(\ref{Eq:mse_loss}) computes fidelity between two $4^n$-dimensional quantum states, incurring prohibitively high computational costs. Secondly, the observable in this function remains neither fixed nor explicit. Due to these limitations, we opt for the following loss function:
\begin{equation}
    \label{Eq:mse_loss_compress}
    \Lc_2 (\Dc, \tht) := 1-\frac{1}{N} \sum_{m = 1}^N [F(\MaxEntangledState_{C_1}, \psi_{C_2})].
\end{equation}
More specifically, 
\begin{equation}
\begin{aligned}
    & \quad F(\MaxEntangledState_{C_1}, \psi_{C_2})  \\
    & = \tr(\MaxEntangledState \tr_{B'}[(\Vc(\theta) \circ U_m \circ \Uc(\theta))(\MaxMixedState_{A'}\ot\MaxEntangledState_{C_1})]).   
\end{aligned}
\end{equation}
In this loss function $\Lc_2$, we only consider the compression process and compute the fidelity between the state on the ``trash" subsystem and the maximally entangled state.
$\Lc_2$ only calculates the fidelity between two $d^2$-dimensional states, while $\Lc_1$ calculate the fidelity between two $D^2$-dimensional states. The observable in $\Lc_2$ $\MaxEntangledState$ that is fixed and explicit.
Moreover, we claim that $\Lc_2 = \Lc_1 = 0$ when the perfect compression is achieved; more detail is discussed in Prop.\ \ref{perfect_compress_recovery_theorem}. While $\Lc_2$ is a more proper loss function, $\Lc_1$ is a proper validation index when we need to evaluate the performance of \VQAname, as it can represent the reconstruction fidelity. 

Furthermore, there is an inevitable issue that the gradient exponential vanishing in a variational quantum algorithm. This issue is called the Barren Plateau (BP) \cite{mcclean2018barren} problem and has been studied in many works, such as Refs. \cite{cerezo2021cost, cerezo2020impact}. In Ref.~\cite{cerezo2021cost}, the relationship between cost function and BP have been discussed. In addition, the authors demonstrated that a local cost function can reduce the adverse effects of BP. 
Inspired by this idea, we design the following loss function to reduce the impact of BP in \VQAname.
\begin{equation}
    \label{Eq:mse_loss_compress_local}
    \begin{aligned}
     &\quad \Lc_3 (\Dc, \tht) \\
     &:=  1-\frac{1}{N} \sum_{m = 1}^N [\tr(O \tr_{B'}[(\Vc(\theta) \circ U_m \circ \Uc(\theta))(\MaxMixedState_{A'}\ot\MaxEntangledState_{C_1})] ))],
    \end{aligned}
\end{equation}
where $O$ is a local observable and 
\begin{equation}
\label{eq: local_observable}
    O = \sum_{k=1}^{n-m} \MaxEntangledState_{k,k+(n-m)} \ot \1_{\overline{k,k+(n-m)}},
\end{equation}
where $n$ and $m$ are the number of system qubits of the original and latent circuits, respectively. Here, $\MaxEntangledState_{k, k+(n-m)}$ is the maximally entangled state on the $k$ and $k+(n-m)$ subsystem, which can be written as 
\begin{equation}
    \label{hamiltonian}
    \begin{aligned}
        \MaxEntangledState_{k, l} &= \frac{1}{2} [\ketb{00}{00} + \ketb{00}{11} + \ketb{11}{00} + \ketb{11}{11}]_{k,l} \\
        &= \frac{1}{2}[I + Z_kZ_{l} + X_{k}X_{l} - Y_{k}Y_{l}].
    \end{aligned}
\end{equation} 


In Ref.~\cite{cerezo2021cost}, the loss function $\Lc_2$, which directly compares the fidelity of two quantum states, is defined as the global cost function; the loss function $\Lc_3$, which is the summation of the expectations of local observables, is defined as the local cost function. The authors prove that the global cost function leads to exponentially vanishing gradient even though the ansatz is shallow and that the local cost function leads to, at worst, polynomially vanishing gradients. 

\begin{figure}[!htbp]
    \centering
    \subfigure[(2,1)]{
    \includegraphics[width=0.2\linewidth]{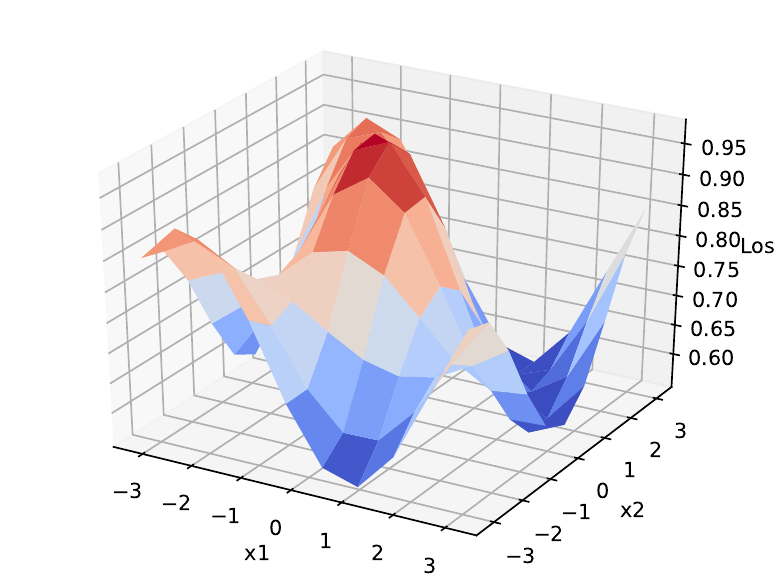}
    }
    \subfigure[(3,1)]{
    \includegraphics[width=0.2\linewidth]{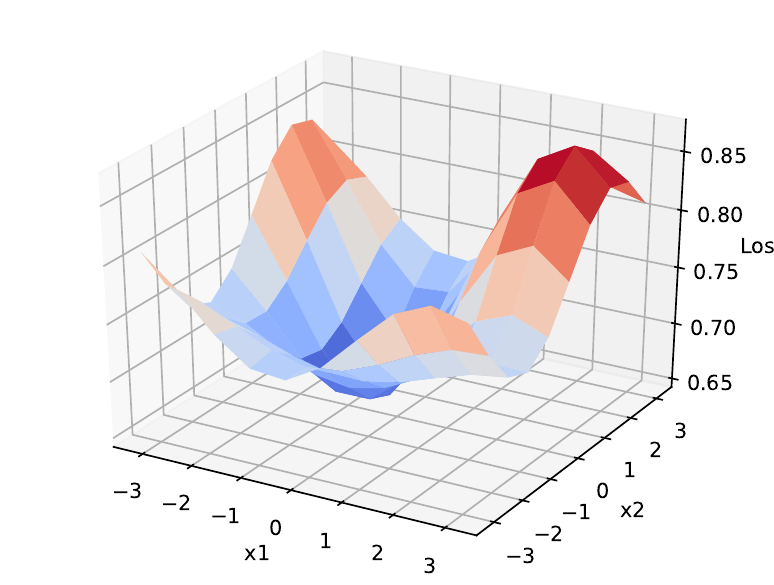}
    }
    \subfigure[(4,1)]{
    \includegraphics[width=0.2\linewidth]{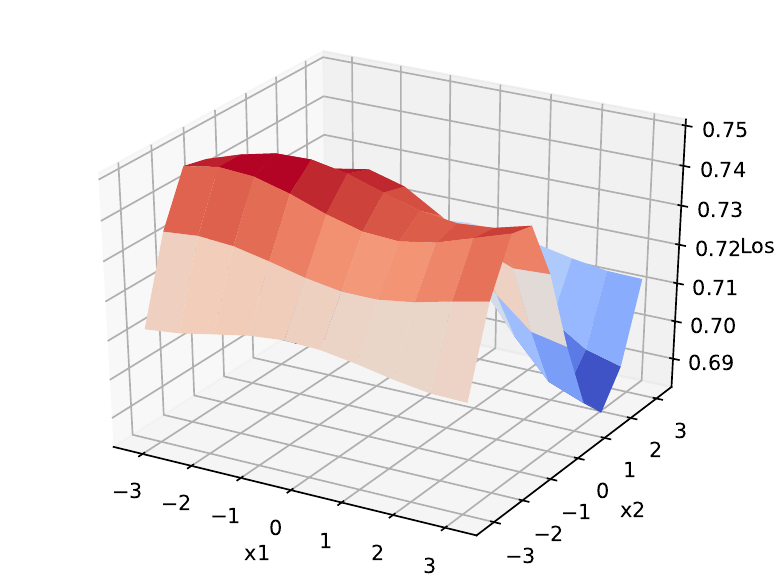}
    }
    \subfigure[(5,1)]{
    \includegraphics[width=0.2\linewidth]{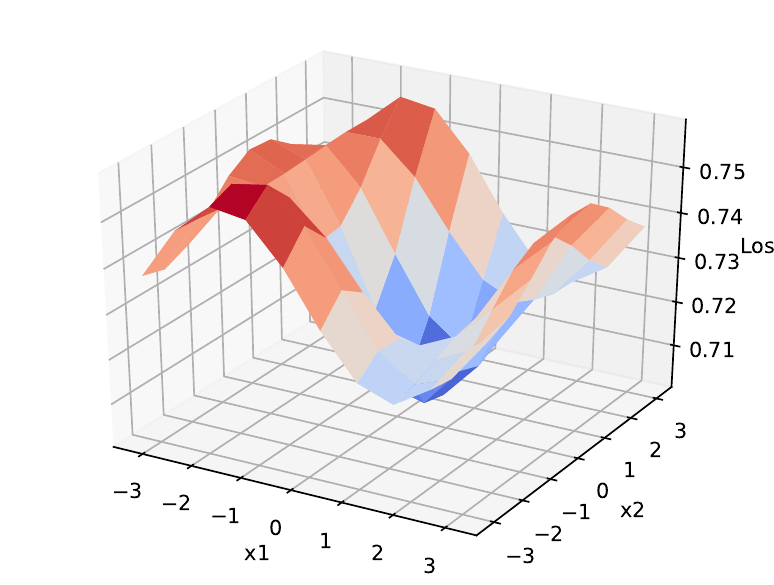}
    }
    \subfigure[(3,2)]{
    \includegraphics[width=0.2\linewidth]{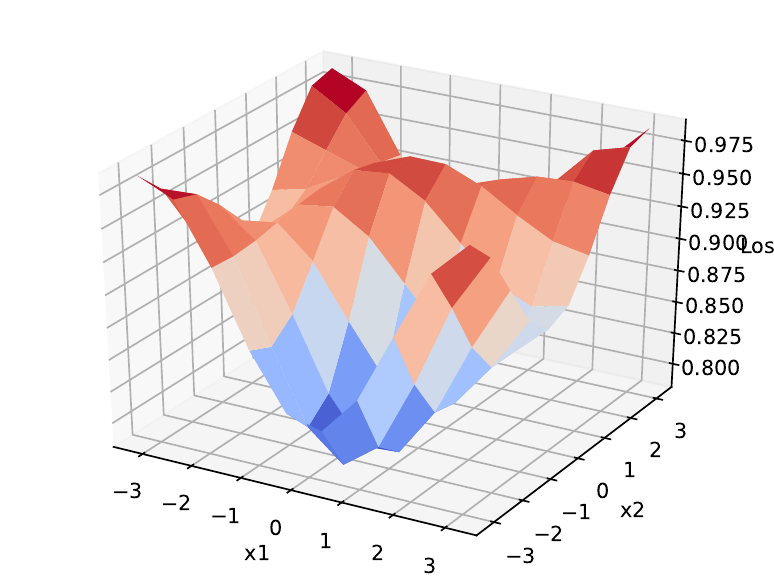}
    }
    \subfigure[(4,2)]{
    \includegraphics[width=0.2\linewidth]{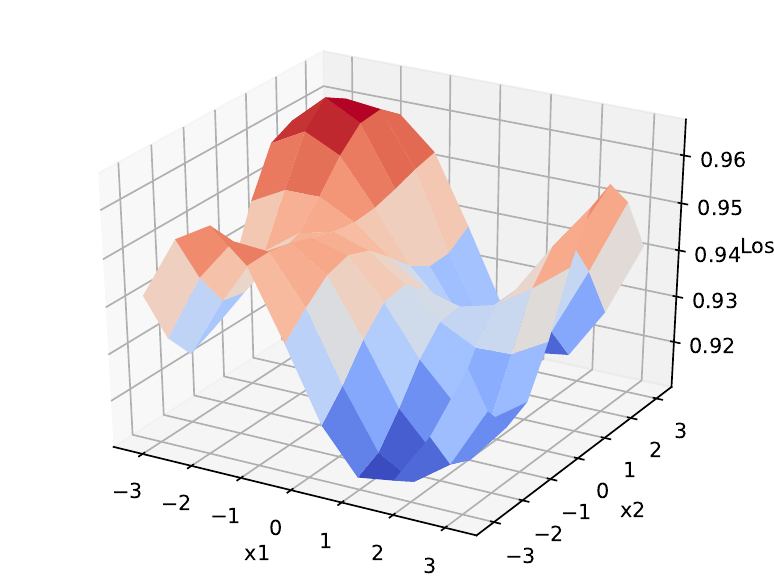}
    }
    \subfigure[(5,2)]{
    \includegraphics[width=0.2\linewidth]{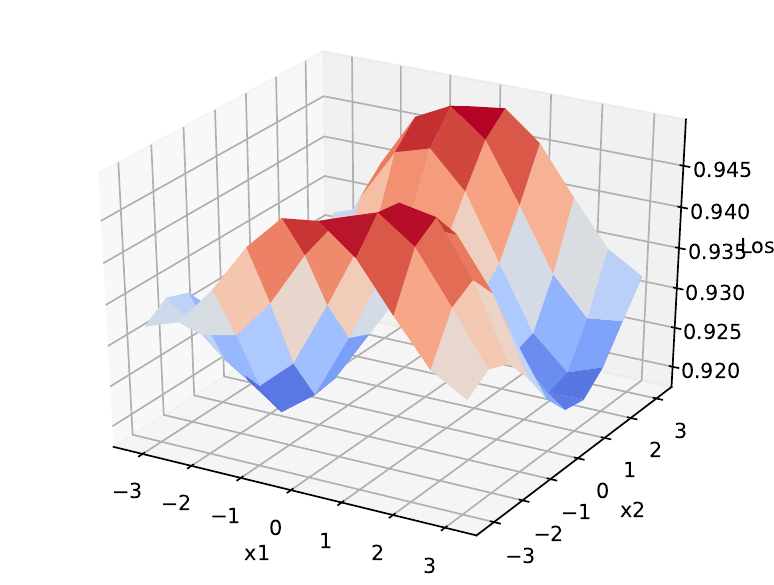}
    }
    \subfigure[(6,2)]{
    \includegraphics[width=0.2\linewidth]{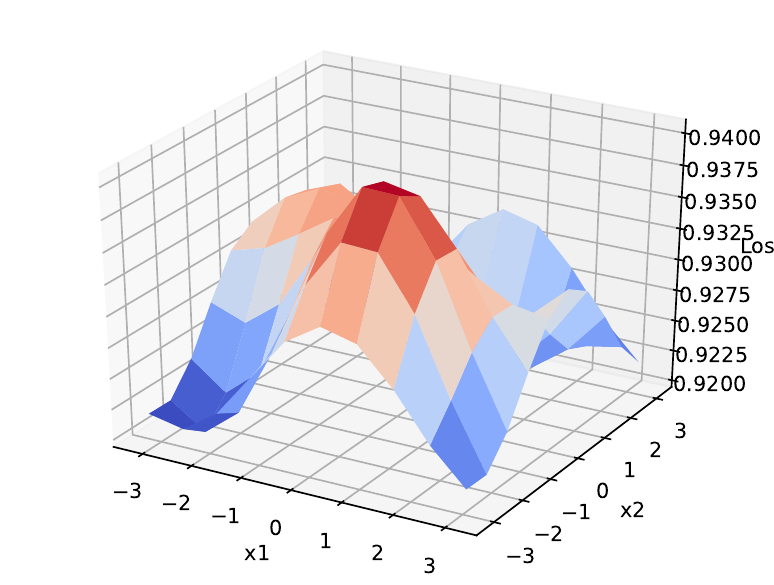}
    }
    \caption{The slice of loss landscape with respect to the first two circuit parameters by changing the input channels' size $n$ and latent channel size $m$. Here, the binary list represents $(n, m)$.}
    \label{fig:landscape}
\end{figure}

In summary, we would like to highlight two critical aspects in the context of our \VQAname~. Firstly, our approach involves performing $(n-m)$-qubit measurements and utilizing the outcomes as the basis for the loss function. Secondly, the observable in the loss function (refer to Eq.~(\ref{Eq:mse_loss_compress_local})) is a summation of several 2-qubit observables.
Consequently, our \VQAname~ may mitigate the barren plateau issue, particularly when the number of layers $L$ is $\Oc(\log(n))$. Fig.~\ref{fig:landscape} provides a visualization of the landscape of \VQAname, depicting a target channel formed by the combination of ten PQCs. 
As illustrated in the figure, the impact of the barren plateau is alleviated across various settings of the [original, latent] qubit pairs.

\subsection{Validation}
\label{sec: validation}
After completing the training process for \VQAname, it is crucial to develop a strategy for evaluating the performance of the training results. During training, our focus is exclusively on the encoding part. This focus must be extended to the entire scheme in the validation process.

After training, we obtained the near-optimal parameters $\tht^\star$ and the reduced channel $\Fc$ from $\Pi$, $\Fc = \tr_{C_1 C_2} [\Vc(\tht^\star) \circ \Ec \circ \Uc(\tht^\star)]$. Finally, $\Tilde{\Ec} = \Vc^\dagger(\tht^\star) \circ (\Fc \ot \id) \circ \Uc^\dagger(\tht^\star)$ is obtained in the decoding process. In the validation part, we will evaluate the similarity between two quantum channels $\Ec$ and $\Tilde{\Ec}$. An simple way is to calculate the fidelity between their Choi states. Two issues must be addressed in the validation process. One is to obtain the reduced channel $\Fc$, and another is to obtain the product channel $\Fc\ot \id$.

To address the first issue, we propose an equivalence problem: extracting the corresponding Choi state $J^{\Fc}$ from $J^{\Pi}$. We prepare the Choi state $J^{\Pi}$ for channel $\Pi$. The reduced Choi state $J^\Fc = \tr_{[m, n],[n+m, 2n]} J^{\Pi}$ is obtained by performing a partial trace on the subsystems from the $m$-th to $n$-th qubits and from the $(n+m)$-th to $2n$-th qubits. As a result, we obtain the reduced Choi state of $J^{\Fc}$ from $\Pi$. 

The second issue is how to efficiently construct the Choi state $J^{\Fc \ot \id}$. Having obtained the reduced Choi state $J^{\Fc}$ from the encoding process, we calculate the Choi state of the identity channel, $J^{\id}$. A direct strategy involves obtaining the state $J^{\Pi} \ot J^{\id}$ and applying the swap operator to adjust the subsystems, yielding $J^{\Fc \ot \id}$.
The swap operator swaps the $[m+1, 2m]$ qubits subsystem with the $[2m+1, m+n]$ qubits subsystem. The issue of designing the circuit consisting of swap circuits is equivalent to a permutation problem.
For this circuit design problem, we propose a strategy, the details of which are presented in Appendix \ref{append: swap_circuit}.

\section{Theoretical Analysis}\label{Sec:Theory}
In this section, we present the key theoretical findings, including the perfect compression condition and the fidelity bound associated with \VQAname. The perfect compression condition can justify our choice of loss function $\Lc_2$. The upper bound on reconstruction fidelity implies that the efficacy of our method is constrained by the rank of the input quantum channel, a parameter intricately tied to the quantity of input quantum circuits. Additionally, the lower bound on reconstruction fidelity, which is under the consideration of the input channel as the depolarizing channel, serves as a performance guarantee for our algorithm.

In the Sec.~\ref{subsection: lossfunction} , we present three distinct loss functions. The preference of $\Lc_2$ over $\Lc_1$ is driven by the objective of downsizing the measurement system from $n$-qubits to $(n-m)$-qubits, a critical step in reducing computational costs. The crucial observation facilitating the transformation of the loss function from $\Lc_1$ to $\Lc_2$ lies in the fact that both $\Lc_1$ and $\Lc_2$ converge to 0 when the input channel can be perfectly recovered after compression. The subsequent proposition provides the analytical insight into this transformation:
\begin{prop}
\label{perfect_compress_recovery_theorem}
\textit{(Perfect compression condition)}
The channel $\Ec$ can be recovered from $\Fc$ by recovery scheme illustrated in Fig.~\ref{fig:QCAE-frame} if and only if
\begin{equation}
\label{Eq: pcc1}
\tr_{B'}\Pi(\MaxMixedState_{A'}\ot\MaxEntangledState_{C_1})=\MaxEntangledState_{C_2},
\end{equation}
where $\MaxEntangledState$ is the maximally entangled state, $\MaxMixedState$ denotes the maximally mixed state, and 
$\Pi = \Vc \circ \Ec \circ \Uc$ is the channel obtained by applying encoders to $\Ec$.
\end{prop}

The proof is shown in Appendix \ref{append: proof_of_perfect_compression}. Prop.~\ref{perfect_compress_recovery_theorem} indicates that the recovery of a quantum channel after compression is feasible if the origin channel can be processed as a product of a compressed channel and an identity channel under the influence of two unitary operators. This proposition implies the feasibility of achieving the learning task, namely finding the optimal $\Uc$ and $\Vc$, by training solely on the ``trash" state.

As an information compression method, it is imperative to assess its performance in terms of recovery. We provide the upper and lower bounds on reconstruction fidelity for \VQAname, with the lower bound derived under the assumption that the input channel is depolarizing.
\begin{lem}
\label{lemma1}
Consider quantum states $\rho$ and $\sigma$, with $r$ being the rank of $\sigma$. The fidelity between $\rho$ and $\sigma$ is bounded above by the sum of the largest $r$ eigenvalues of $\rho$. This bound is attained if and only if $\rho=\sigma$.
\end{lem}

Lem.~\ref{lemma1} provides an upper bound on the fidelity between any two quantum states, a result instrumental in proving the following proposition.

\Prop{
\label{recovery_fidelity_bounded}
Consider $\Tilde{\Ec}$ as the recovered quantum channel from $\Vc\circ\Ec\circ\Uc$, the recovery fidelity $F(\Tilde{\Ec}, \Ec)$ is bounded above by the sum of the largest $d^2$ eigenvalues of the Choi state of $\Ec$, where $d$ is the dimension of the reduced quantum channel  $\Fc = \tr_{trash}[\Vc\circ\Ec\circ\Uc]$.
}

The proofs for these two results are detailed in Appendix \ref{append: proof_of_upper_bound}. Drawing inspiration from this proposition, it becomes evident that the reconstruction fidelity via \VQAname~may is not optimal when the rank exceeds $d^2$. For instance, consider the completely depolarizing quantum channel $\Delta$ with input and output dimensions $D$ and compress it to $d$ dimensions. The Choi state of $\Delta$ is $\textnormal{diag}(\frac{1}{D^2}, \cdots, \frac{1}{D^2})$. According to the proposition mentioned above, even under the best-case training scenario, the fidelity of reconstruction remains bounded by $\frac{d^2}{D^2}$.

While upper bounds provide valuable insights, lower bounds are also important as it offers a performance guarantee, at least for this special case. In this study, we explore the lower bound of reconstruction fidelity when the input channel is the depolarizing channel, as outlined in the following proposition.

\Prop{
\label{lower_bonud_for_QCAE_on_dep}
For a given depolarizing channel $\Ec_p$ with dimension $D$, utilizing \VQAname~to compress it to a $d$-dimension quantum channel and subsequantly recover it to $D$-dimension, the lower bound on the reconstruction fidelity is given by 
\begin{equation}
    F(\Ec_p, \Tilde{\Ec_p}) \geqslant \left[\sqrt{(\frac{p}{D^2} + 1 - p)(\frac{p}{d^2} + 1 - p)} + (d^2-1)\frac{p}{Dd}\right]^2.
\end{equation}
}
The proof is presented in Appendix \ref{append: proof_of_lower_bound}. 

\section{Applications and Numerical Experiments\label{app_and_evaluate}}
In this section, we delve into practical applications of \VQAname, with a primary focus on quantum circuit information compression, the fundamental objective motivating the proposal of \VQAname. Additionally, we explore two other applications: \VQAname-based anomaly detection and denoising for quantum circuits. 
In our experiments, we applied compression and reconstruction techniques to multiple quantum circuits, achieving remarkably low reconstruction error rates, with approximately 0.05. Moreover, \VQAname~ has demonstrated remarkable efficacy in detecting ``abnormal" data from ``normal" data and mitigating the noise on quantum circuits.

We utilized the quantum platforms Qiskit and Mindquantum in our experiments. The updated code is available at \cite{wjqcae}. 

\subsection{Quantum Circuit Information Compression}
This section investigates the capability of leveraging \VQAname~for information compression on quantum circuits. As illustrated in Sec.~\ref{Sec:method}, we consider a set of quantum circuits $\{ U_i \}_{i=1}^{N}$ with a dimension $D$, constructing these circuits as a mixed quantum channel $\Ec$. The encoding process involves finding a supermap to map the $D$-dimensional quantum channel $\Ec$ to a $d$-dimensional channel $\Fc$. The encoding process compresses the information within the quantum circuits. A critical metric for evaluating compression performance is the reconstruction fidelity between the original channel $\Ec$ and the reconstructed channel $\Tilde{\Ec}$ which is recovered from $\Fc$.

Our experiments focused on compressing information within parameterized quantum circuits (PQCs). PQCs serve as widely-used encoding tools for translating classical information into quantum information in Quantum Neural Networks (QNNs), owing to their solid expressive power. Specifically, We target the RealAmplitudes from the qiskit circuit library as the PQCs for compression. The parameters required for these circuits are independently generated using a normal distribution.

\begin{figure}[!htbp]
    \centering
    \begin{minipage}[c]{0.38\textwidth}
        \centering
        \includegraphics[width=\linewidth]{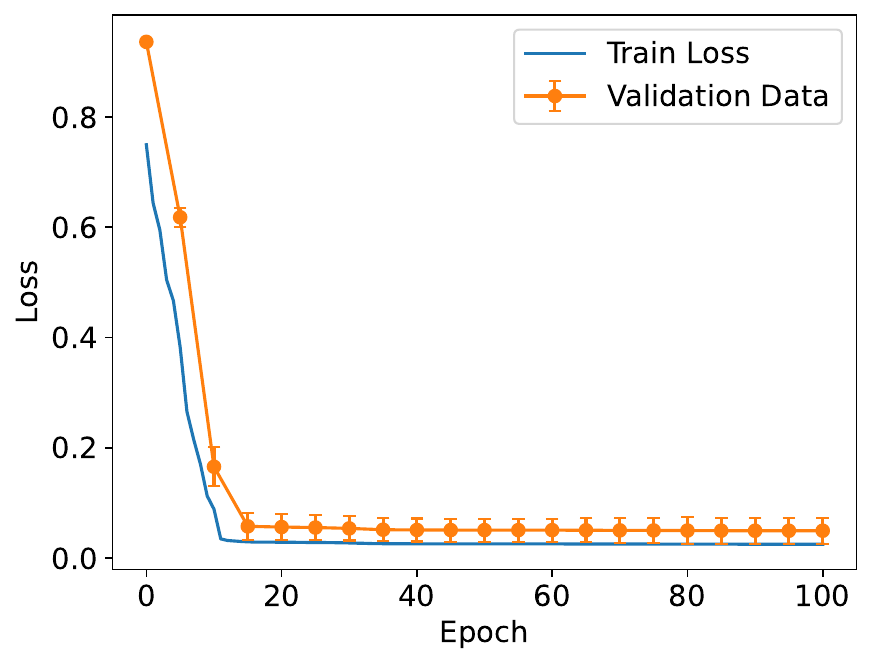}
    \end{minipage}
    \caption{The training process of \VQAname~ for circuit information compression. The PQC used is the RealAmplitudes circuits for encoding classical information to quantum states, and the parameters to control the pqcs are generated with the normal distribution $N(0, 0.1)$. The number of the circuits is 50, and the training epoch is set as 100.}
    \label{fig:compress_performance}
\end{figure}

Fig.~\ref{fig:compress_performance} showcases the experimental results. In this experiment, we utilize \VQAname~ to compress 50 parameterized 4-qubits quantum circuits with the RealAmplitudes construction to 3-qubits cirucits. The control parameters are generated from a normal distribution $N(0, 0.1)$. 
We select L-BFGS-B as the classical optimizer, set the training epochs to 100. The change in the loss function value during the training process is the blue line. The orange line represents each epoch's average validation infidelity and its standard deviation.  
The experiment reveals that the loss converges rapidly in 20 epochs, achieving a reconstruction error of approximately 0.05 for 50 quantum circuits.

The sensitive analysis of super parameters are shown in Appendix~\ref{Sec:sensitive analysis}. 

\subsection{Anomaly Detection}
In this section, we apply \VQAname~ to identify anomalies in quantum circuits. Considering the scenario of chip anomaly detection, the objective is to identify abnormal chips within a collection of quantum chips. Classical data anomaly detection method may not be seamlessly applicable in this scenario. The \VQAname, leveraging variational algorithms, offers a solution tailored to the intricacies of quantum circuit data.

Conventional autoencoder can be used for anomaly detection\cite{an2015variational}. 
The autoencoder learns the encoding distribution of ``normal" data, so when a data set is provided to the autoencoder, it encodes and decodes following the encoding distribution of ``normal" data. ``Normal" data that follows this distribution can achieve lower reconstruction errors. In comparison, anomalous data that does not conform to the distribution will result in higher reconstruction errors.

The specific framework is:
The input is the ``normal" dataset $X$, anomalous dataset $\{x^{(i)}, i = 1, \cdots, N\}$ and a threshold $\alpha$. Then, design an autoencoder network and train it using the ``normal" dataset $X$. Next, for each data $x^{(i)}$ in the anomalous dataset, we use the trained autoencoder to obtain the reconstruction $error(i)$. Finally, make the decision, label the $i$-th data $x^{(i)}$ as ``abnormal" if $error(i) > \alpha$ and ``normal" otherwise.   
\begin{figure}[!htbp]
    \centering
    \subfigure[$N(0,0.1)$ v.s. random circuits]{
    \label{fig:norm_random}
    \includegraphics[width=0.4\textwidth]{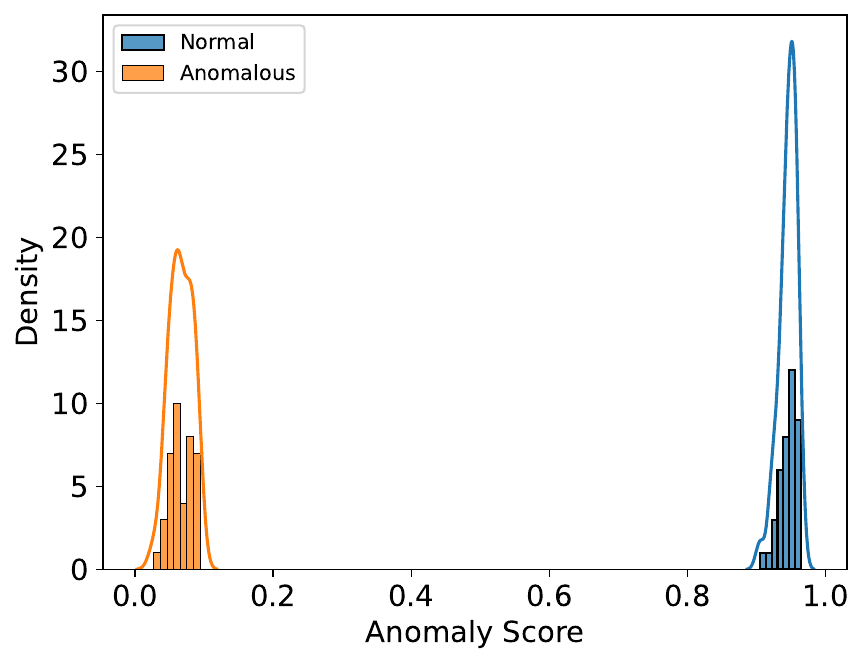}
    }
    \hfill
    \subfigure[$N(0,0.1)$ v.s. $N(5,0.1)$]{
    \label{fig:two_norm}
    \includegraphics[width=0.4\textwidth]{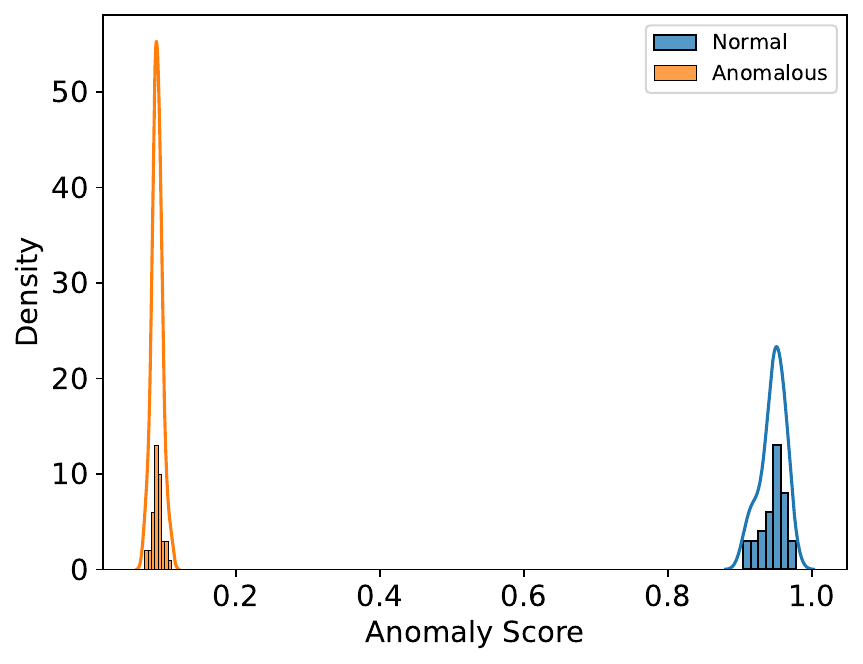}
    }
    \caption{The results of quantum circuits anomaly detection. Both the two figures are the anomaly scores distributions. In Fig.~\ref{fig:norm_random}, RealAmplitudes pqcs with parameters generated by $N(0, 0.1)$ are set as ``normal" data and random quantum circuits are set as ``abnormal" data. In Fig.~\ref{fig:two_norm}, RealAmplitudes pqcs with parameters generated by $N(0, 0.1)$ are also set as ``normal" data, RealAmplitudes pqcs with parameters generated by $N(5, 0.1)$ are set as ``abnormal" data. In experiments, we randomly choose 10 ``normal" circuits to train \VQAname and choose 40 ``normal" circuits in the rest of the ``normal" dataset and 40 ``abnormal" circuits for validation. The blue hists show the anomaly scores of the ``normal" test dataset, and the orange hists show the anomaly scores of the ``abnormal" test dataset. The x label is the reconstruction fidelities and is set as the anomaly scores, and the y label is the density of the circuits with the same anomaly scores.}
    \label{fig:anomaly_detection}
\end{figure}

Similar to the conventional autoencoder, we investigate applying \VQAname~ to detect anomalous quantum circuit tasks. For the given ``normal'' quantum circuits set $\{U_{normal}\}$ and anomalous quantum circuits set, we train a quantum circuit autoencoder using $\{U_{normal} \}$. We also use the reconstruction fidelity as anomalous scores for each circuit in the anomalous quantum circuits set. If the reconstruction fidelity is bigger than a given threshold, we label this circuit as ``normal". Otherwise, we label it as ``abnormal".

The following experiment demonstrates the potential of \VQAname-based quantum circuit anomaly detection. 

{\em Data preparation:} We prepare a quantum circuit dataset based on the RealAmplitudes parameterized quantum circuit. The difference between ``normal" and ``abnormal" circuits is the control parameters in parameterized quantum circuits. The control parameters of two different datasets conform to two different normal distributions. We also generate the random quantum circuits as another ``abnormal" dataset. 

The results are shown in Fig.~\ref{fig:anomaly_detection}, provide that \VQAname~ can be highly effective in detecting abnormal data from normal data. Blue hists and line show the distribution of the anomaly scores of the ``normal" data and orange hists and line show the distribution of the anomaly scores of the ``abnormal" data. In Fig.~\ref{fig:norm_random}, the PQCs whose parameters are distributed as $N(0,0.1)$ as the ``normal" data, and the random circuits generated by qiskit with depth ten as the ``abnormal" data. In Fig.~\ref{fig:two_norm}, the ``normal" data is same and the PQCs whose parameters are distributed as $N(5,0.1)$ is defined as the ``abnormal" data. All circuits are 4 qubits, and the train data consists of 10 ``normal" circuits; the test data consists of 40 ``normal" and 40 ``abnormal" circuits. We use the L-BFGS-B optimizer and set the training epoch as 100.

\subsection{Quantum Circuit Denoise}
In this section, we consider applying \VQAname~ to denoise quantum circuits. In the NISQ era, circuit execution is limited by the effect of noise. An essential application of the conventional autoencoder is denoising data. The main idea is to extract the main character of data by autoencoder under the assumption that the noise in data is not the main feature. Ref.~\cite{bondarenko2020quantum} also proposed a strategy to use quantum autoencoder denoise spin-flip errors and random unitary transformation errors concerning the GHZ state. In this work, we consider denoising the depolarizing error on quantum circuits.

{\em Depolarizing error:}
For an $n$-qubit quantum state $\rho$, the depolarizing channel error $\Ec_p$ affect $\rho$ according to
\begin{equation}
    \begin{aligned}
        \Ec_p(\rho) &= (1-p) \rho \\
                    &+ \sum_{\sigma^{k} \in \{ I, X, Y, Z\}} \frac{p}{2^n}  \bigg(\bigotimes_{i=1}^{n} \sigma_i^k\bigg) \rho \bigg(\bigotimes_{i=1}^{n} \sigma_i^k\bigg),
    \end{aligned}
\end{equation}
where $p$ is the probability of being replaced, and $\sigma_i^k \in \{I, X, Y, Z\}$ is the pauli operator acting on the $i$-th qubit.  

For a quantum circuit $U$, $U$ is afffect by the depolarizing channel error $\Ec_p$ by
\begin{equation}
\label{eq: depo_on_circuit}
    \begin{aligned}
        U \circ \Ec_p(\rho) &= (1-p) U\rho U^\dagger \\
        &+ \sum_{\sigma^{k} \in \{ I, X, Y, Z\}} \frac{p}{2^n}  \bigg(\bigotimes_{i=1}^{n} \sigma_i^k\bigg) U\rho U^\dagger \bigg(\bigotimes_{i=1}^{n} \sigma_i^k\bigg). 
    \end{aligned}
\end{equation}

\begin{figure}
    \centering
    \includegraphics[width = 0.4\textwidth]{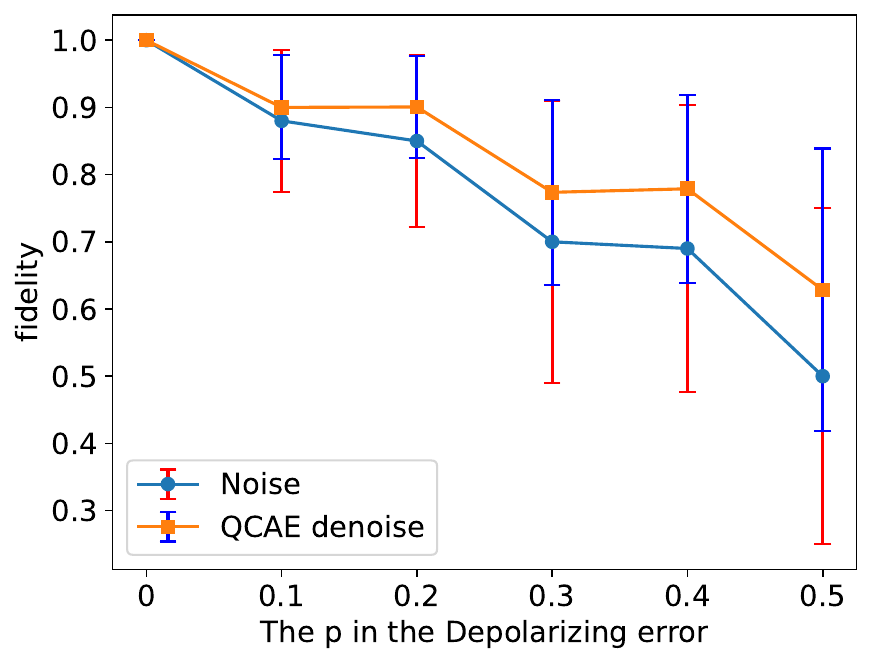}
    \caption{Quantum Circuit AutoEncoder denoising the parameterized quantum circuit under the depolarizing error. We show the average fidelity of noisy test states with the 3-qubit amplitude PQC with parameters under the norm distribution norm-(0, 0.6) before denoising (blue dots) and after denoising (orange squares). Error bars display standard deviations. 100 noisy training pairs, 100 training rounds and the L-BFGS-B optimizer. The number of qubits in original and latent circuits is set as 3 and 1.}
    \label{fig:depolar_denoise_impact}
\end{figure}

The main components for quantum circuit denoising are as follows:

{\em Data preparation}: For a given quantum circuit $U$, we sample a set ${U_i}$ from $U$ under the depolarizing channel $\Ec_p$. More specifically, Eq.~(\ref{eq: depo_on_circuit}) shows that the depolarizing channel is a weighted operator summation; each weight is the probability of adding the operator to the input circuit. In our experiment, we sample an operator $O_i$ with probability and set $U_i = O_i \circ U$. We can obtain the training set $\{U_i\}$ by repeating this process. 

{\em Denoising based on \VQAname}: In this step, we use the training dataset $\{U_i\}$ to train a \VQAname~model. We have described this process in detail in this paper and will not repeat it here. After the training, we can use the \VQAname~model to obtain the reconstruction data set $\{ \Tilde{U_i} \}$.

{\em Validation}: A key issue is evaluating the performance of the circuit denoise. We compute two indices for evaluation. 
One index is the sample impact, which reflects the similarity between the training set $\{U_i\}$ and $U$. We give the mean and variance of the similarities. Another index is the reconstruction impact or denoise performance, which reflects the similarity between the training set $\{U_i\}$ and $U$. We also calculate the mean and variance values.

Fig.~\ref{fig:depolar_denoise_impact} is the result of denoising the parameterized quantum circuit under the depolarizing error. The original quantum circuit is the RealAmplitude circuit with the parameters generated by the distribution $N(0,0.6)$. We sample 100 noise circuits and train the 100 epochs by \VQAname. The number of qubits of original circuits is 3 and 1 for the latent circuits. The blue dots are the mean fidelity of sample data. The orange squares are the mean fidelity after denoise. As a conclusion, the \VQAname~can mitigate the noise impact.

%
\section{Conclusion and Discussion \label{conclu}}
In this work, we introduce the Quantum Circuit Autoencoder model and design a variational quantum algorithm for its implementation, referred to as \VQAname.  Subsequently, we proposed the theoretical analysis to determine the condition for faithful compression, aiding in constructing the local loss function of \VQAname.  Additionally, we establish an upper bound on the reconstruction fidelity of \VQAname, and calculate the fidelity lower bound for cases involving the depolarizing channel as the input channel.  Moreover, we demonstrated the application of \VQAname~ in various toy scenarios, such as information compression, anomaly detection, and denoising for quantum circuits.  Finally, we performed numerical evaluations and implemented \VQAname~ applications using the Qiskit and Mindquantum platforms.

There is much potential for further progress. (1) Determining tasks suitable for \VQAname. On the one hand, quantum circuit autoencoders have applications in data generation and feature extraction for information within quantum circuits. On the other hand, investigating practical applications rather than toy experiments in this work is also crucial work. In addition, finding more practical tasks beyond anomaly detection using Parameterized Quantum Circuits (PQCs) with different parameters and distributions is also appealing. (2) The reconstruction fidelity of \VQAname~is bounded in Prop.~\ref{recovery_fidelity_bounded}. Using the noise-assisted channel to overcome the fidelity limited in \VQAname~ as discussed in Ref.~\cite{cao2021noise}, which uses a noise-assisted channel to overcome the fidelity limited in QAE. (3) In \VQAname, we use the PQCs as the encoders and decoders in \QCAEmodel. It might be more powerful to substitute the PQCs with the parameterized quantum channels. (4) This work only considers the lower bound on reconstruction fidelity for special cases, and it would be an interesting question to consider the general cases.

\begin{acknowledgments}
This work was partially supported by the National Natural Science Foundation of China (Grant No.~62102388), 
Innovation Program for Quantum Science and Technology (Grant No.~2021ZD0302900), 
and Anhui Initiative in Quantum Information Technologies (Grant No.~AHY150100).
\end{acknowledgments}
\appendix

\setcounter{thm}{0}

\section{Proof of \textit{The Perfect Compression Condition}}
\label{append: proof_of_perfect_compression}
\Prop{
\label{append-perfect_compress_recovery_theorem}
\textit{(Perfect compression condition)}
The channel $\Ec$ can be recovered from $\Fc$ by recovery scheme illustrated in Fig.~\ref{fig:QCAE-frame} if and only if
\begin{equation}
\label{append-Eq: pcc1}
\tr_{B'}\Pi(\om_{A'}\ot\MaxEntangledState_{C_1})=\MaxEntangledState_{C_2},
\end{equation}
where $\MaxEntangledState$ is the maximally entangled state, $\om$ denotes the maximally mixed state, and $\Pi = \Vc \circ \Ec \circ \Uc$ is the channel obtained by  applying encoders to $\Ec$.}
\begin{proof}
\textit{The condition is sufficient:} If $\Ec$ can be recovered from $\Fc$ by the decoding scheme in Fig.~\ref{fig:QAE-frame} faithfully, we can get 
\begin{equation}
    \Ec = \Vc^\dagger \circ (\Fc \ot \1) \circ \Uc^\dagger, 
\end{equation}
and this means 
\begin{equation}
    \Pi = \Vc \circ \Ec \circ \Vc = \Fc_{A' \to B'} \ot \1_{C_1 \to C_2},
\end{equation}
which means that the channel $\Pi$ is a product channel, and the sub-channel on the subsystem $C_1$ is identity. That is,  $\MaxEntangledState_{C_2} = \MaxEntangledState$.

\textit{The condition is necessary:} If Eq.~(\ref{Eq: pcc1}) is satisfied.
Let $J^\Pi_{A' C_ 1 B' C_2}$ be the Choi state of $\Pi_{A' C_1\to B' C_2}$, we can deduce from Eq.~(\ref{evolution_of_choi}), the result state after apply $\Pi$ to initial state $\om_{A'} \ot \psi_{C_1}$:
\begin{equation}
    \begin{aligned}
    &\Pi_{A' C_1\to B' C_2}(\om_{A'} \ot \psi_{C_1}) \\
    &= D\tr_{A' C_1} J^\Pi_{A' C_1 B' C_2} (\om_{A'} \ot \MaxEntangledState_{C_1} \ot \1_{B' C_2}).
    \end{aligned}
\end{equation}
Since $\tr_{B'}\Pi(\om_{A'}\ot\MaxEntangledState_{C_1})=\MaxEntangledState_{C_2}$, we have 
\begin{equation}
    \label{append-Eq: 15}
    \begin{aligned}
    &\tr_{B'}\Pi(\om_{A'}\ot\MaxEntangledState_{C_1}) \\
    &= D\tr_{A' C_1B'}(J^\Pi_{A' C_1 B' C_2}(\om_{A'}\ot\MaxEntangledState_{C_1}\ot\1_{B' C_2})) \\
    &= \MaxEntangledState_{C_2}.
    \end{aligned}    
\end{equation}
Since
\begin{equation}
    \tr[M_{AB}(\rho_A \ot \1_{B})] = \tr [(\tr_{B} M_{AB}) \rho_A],
\end{equation}
where $M_{AB}$ is a quantum operation (or channel) on a composite system $\Hc_A \ot \Hc_B$ and $\rho_{A}$ is a density operator on $\Hc_A$,we have
\begin{equation}
    \label{append-Eq: 17}
    \begin{aligned}
    \MaxEntangledState_{C_2} &= D\tr_{A' C_1B'}(J^\Pi_{A' C_1 B' C_2}(\om_{A'}\ot\MaxEntangledState_{C_1}\ot\1_{B' C_2}))\\
    &= \frac{D}{d}\tr_{C_1}(J^\Pi_{C_1 C_2}(\MaxEntangledState_{C_1}\ot\1_{C_2})) = \Pi_{C_1 \to C_2}(\MaxEntangledState_{C_1}), 
    \end{aligned}
\end{equation}
which implies that the reduced quantum channel $\Pi_{C_1\to C_2}$ of $\Pi$ is an identity channel, so the quantum channel $\Fc$ can be deemed a $d$-dimensional channel with input system $A'$ and output system $B'$.

As the state $\MaxEntangledState_{C_2}$ is a maximally entangled state, so the state after apply $\Pi$ to initial state $\om_{A'} \ot \MaxEntangledState_{C_1}$ is product state $\om_{B'} \ot \MaxEntangledState_{C_2}$. This means $\Pi$ can be written as the form of $\Fc \ot \1$, which means that we can recover quantum channel by:
\begin{equation}
    \label{append-Eq: 18}
    \Vc^{\dagger}\circ (\Fc \ot \1)  \circ \Uc^{\dagger} = \Vc^{\dagger}\circ \Pi \circ \Uc^{\dagger} = \Vc^{\dagger}\circ \Vc\circ\Ec\circ\Uc \circ \Uc^{\dagger} = \Tilde{\Ec},
\end{equation}
as depicted in the decoding process of Fig.~\ref{fig:QCAE-frame}.
\end{proof}
The inductions from Eq.~(\ref{append-Eq: 15}) to Eq.~(\ref{append-Eq: 17}) and form Eq.~(\ref{append-Eq: 17}) to Eq.~(\ref{append-Eq: 18}) use the fact that the initial state is a product state of $\MaxMixedState$ and $\MaxEntangledState$.

\section{Proof of \textit{The Upper Bound of \VQAname}}
\label{append: proof_of_upper_bound}
\begin{lem}
\label{append-lemma1}
Consider quantum states $\rho$ and $\sigma$, with $r$ being the rank of $\sigma$. The fidelity between $\rho$ and $\sigma$ is bounded above by the sum of the largest $r$ eigenvalues of $\rho$. This bound is attained if and only if $\rho=\sigma$.
\end{lem}
\begin{proof}
\begin{equation}
\label{append-dfkjbwdgb}
    \begin{aligned}
    F(\rho, \sigma) &= F(\Pi_{\im(\sigma)} \rho \Pi_{\im(\sigma)}, \sigma)\\
                    &\leqslant \tr(\Pi_{\im(\sigma)} \rho \Pi_{\im(\sigma)}) \tr (\sigma) \\
                    &= \tr(\Pi_{\im(\sigma)} \rho) 
                    = \sum_{i=1}^{r} \lambda_i,
    \end{aligned}
\end{equation}
where $\Pi_{\im(\sigma)}$ denote the projection onto the image of $\sigma$, $\lambda = (\lambda_1,  \cdots,\lambda_r)$ is the eigenvalues of $\rho$ and $\lambda_1 \geqslant  \cdots \geqslant \lambda_r$. And the inequality in \eqref{append-dfkjbwdgb} is due to the Proposition 3.12 in \cite{watrous2018theory}.
\end{proof}
Lem.~\ref{lemma1} give us a upper bound on the fidelity between any two quantum state, and it can be used to prove following proposition. 


\Prop{
\label{append-recovery_fidelity_bounded}
Consider $\Tilde{\Ec}$ as the recovered quantum channel from $\Vc\circ\Ec\circ\Uc$, the recovery fidelity $F(\Tilde{\Ec}, \Ec)$ is bounded above by the sum of the largest $d^2$ eigenvalues of the Choi state of $\Ec$, where $d$ is the dimension of the reduced quantum channel  $\Fc = \tr_{trash}[\Vc\circ\Ec\circ\Uc]$.
}
\begin{proof}
Let $J_{A'B'}^\Fc$ and $J_{A' C_1 B' C_2}^\Pi$ be the Choi state of $\Fc$ and $\Ec$, respectively. 
\begin{equation}
    \begin{aligned}
        F(\Tilde{\Ec}, \Ec)
        &= F(\Vc^{\dagger}\circ ((\Vc\circ\Ec\circ\Uc)_{A' \to B'} \ot \id_{C_1 \to C_2}) \circ \Uc^{\dagger}, \Ec)\\
        &= F((\Vc\circ\Ec\circ\Uc)_{A' \to B'} \ot \id_{C_1 \to C_2}, \Vc \circ \Ec \circ \Uc)\\
        &= F(J_{A'B'}^\Fc \ot \MaxEntangledState_{C1,C2}, J_{A' C_1 B' C_2}^\Pi).
    \end{aligned}
\label{append-Eq: full_fidelity}
\end{equation}
It is easy to show that
\begin{equation}
    rank(J_{A' B'}^\Fc \ot \MaxEntangledState_{C_1,C_2}) \leqslant d^2.
\label{append-Eq: rank}
\end{equation}
And by Lem.~\ref{lemma1},  we can get
\begin{equation}
    \begin{aligned}
        &F(J_{A'B'}^\Fc \ot \MaxEntangledState_{C1,C2}, J_{A' C_1 B' C_2}^\Pi) 
        \leqslant \sum_{i=1}^{d^2} \lambda_i = \sum_{i=1}^{d^2} \mu_i,
    \end{aligned}
\end{equation}
 where $\lambda = (\lambda_1, \lambda_2, \cdots, \lambda_{d^2})$ is the eigenvalues of $J^\Fc$ with $\lambda_1 \geqslant \lambda_2 \geqslant \cdots \geqslant \lambda_{d^2}$, $\mu = (\mu_1, \mu_2, \cdots, \mu_{d^2})$ is the eigenvalues of $J^\Ec$ and $\mu_1 \geqslant \mu_2 \geqslant \cdots \geqslant \mu_{d^2}$.
\end{proof}




\section{Proof of \textit{The fidelity lower bound of \VQAname on compress depolarizing channel}}
\label{append: proof_of_lower_bound}
\Prop{
\label{append-lower_bonud_for_QCAE_on_dep}
For a given depelorizing channel $\Ec_p$ with dimesion $D$, using the \VQAname~to compress it to a $d$-dimension quantum channel, and recovery it to $D$-dimension, the lower bound of the reconstruction fidelity is 
\begin{equation}
    F(\Ec_p, \Tilde{\Ec_p}) \geqslant \left[\sqrt{(\frac{p}{D^2} + 1 - p)(\frac{p}{d^2} + 1 - p)} + (d^2-1)\frac{p}{Dd}\right]^2.
\end{equation}
}
\begin{proof}
For an arbitrary quantumm channel $\Ec$, let $\Pi_{A'C_1 \to B'C_2} := \Vc \circ \Ec \circ \Uc$ and $\Fc_{A' \to B'} = \tr_{C_1 C_2}\Pi$, the recovery channel $\Tilde{\Ec} = \Vc^\dagger \circ (\Fc_{A' \to B'} \ot \id_{C_1 \to C_2}) \circ \Uc^\dagger$. The states $J_{AB}^{\Ec}$, $J_{A'B'}^\Fc$, $J_{A' C_1 B' C_2}^\Pi$ and $J_{AB}^{\Tilde{\Ec}}$ are the Choi states of $\Ec$, $\Fc$, $\Ec$ and $\Tilde{\Ec}$, respectively. 

Define the reconstruction fidelity as 
\begin{equation}
\begin{aligned}
    F(\Ec, \Tilde{\Ec}) &= \max_{\Uc, \Vc} F(\Ec, \Vc^\dagger \circ (\Fc_{A' \to B'} \ot \id_{C_1 \to C_2}) \circ \Uc^\dagger) \\
    &= \max_{\Uc, \Vc} F(\Vc \circ \Ec \circ \Uc, \Fc_{A' \to B'} \ot \id_{C_1 \to C_2}) \\
    &= \max_{\Uc, \Vc} F(\Pi_{A'C_1 \to B'C_2}, \Fc_{A' \to B'} \ot \id_{C_1 \to C_2})\\
    &= \max_{\Uc, \Vc} F(J^{\Pi}_{A'C_1 \to B'C_2}, J^{\Fc}_{A' \to B'}\ot \MaxEntangledState_{C_1 C_2}).
\end{aligned}
\end{equation}
Setting $\Uc=\Vc=\id$ yields a lower bound as follows.
\begin{equation}
\begin{aligned}
\label{eq: lower_bound_inequality}
    F(\Ec, \Tilde{\Ec}) &\geqslant  F_{\{\Uc = \Vc = \id\}}(J^{\Pi}_{A'C_1 \to B'C_2}, J^{\Fc}_{A' \to B'}\ot \MaxEntangledState_{C_1 C_2}), 
\end{aligned}
\end{equation}
Eq.~(\ref{eq: lower_bound_inequality}) means that the reconstruction fidelity when the encoders and decoders are all is identity is a lower bound.

For the given depolarizing channel $\Ec_p$, 
\begin{equation}
    J^{\Ec_p} = p\MaxMixedState_{D\times D} + (1-p)\MaxEntangledState_{D},
\end{equation}
where $\MaxMixedState_{DxD} = \frac{\1}{D^2}$ is the maximally entangled state and $\MaxEntangledState_{D} = \sum_{i,j=0}^{D-1} \ketb{i}{j}\ot\ketb{i}{j}$. 
\begin{equation}
\label{eq: depo_fidelity_function}
\begin{aligned}
    &F_{\{\Uc = \Vc = \id\}}(J^{\Ec_p}_{A'C_1 \to B'C_2}, J^{\Fc}_{A' \to B'}\ot \MaxEntangledState_{C_1 C_2})\\
    &= F(p\MaxMixedState_{D\times D} + (1-p)\MaxEntangledState_{D}, p \frac{\1_{d\times d}}{d^2} \ot \MaxEntangledState_{D/d} + (1-p)\MaxEntangledState_{D}).
\end{aligned}
\end{equation}

Let $\ket{\psi_1}, \cdots, \ket{\psi_{D^2}}$ be an orthogonal basis of the $D^2$ Hilbert space. The basis satisfy that
\begin{equation}
    \MaxEntangledState_D = \ketb{\psi_1}{\psi_1} = \MaxEntangledState_d \ot \MaxEntangledState_{D/d} = \ketb{\psi_1'}{\psi_1''} \ot \ketb{\psi_1'}{\psi_1''}, 
\end{equation}
and
\begin{equation}
\begin{aligned}
    \ket{\psi_i} = \ket{\psi_i'} \ot \ket{\psi_1''},\quad i\in [ 2, \cdots, d^2],
\end{aligned}
\end{equation}
where $\{ \ket{\psi_i'} \}$ is an orthogonal basis of the $d^2$ Hilbert space.

The spectral decomposition of the two quantum state in the fidelity function in Eq.~(\ref{eq: depo_fidelity_function}) is 
\begin{equation}
\begin{aligned}    
    & p\MaxMixedState_{D\times D} + (1-p)\MaxEntangledState_{D} = \sum_{i=1}^{D^2} \lambda_i \ketb{\psi_i}{\psi_i},\\
    & \lambda_1 = \frac{p}{D^2} + 1 - p, \\
    & \lambda_2 = \cdots = \lambda_{D^2} = \frac{p}{D^2},
\end{aligned}
\end{equation}
and 
\begin{equation}
\begin{aligned}
     & p \frac{\1_{d\times d}}{d^2} \ot \MaxEntangledState_{D/d} + (1-p)\MaxEntangledState_{D} = \sum_{i=1}^{D^2} \mu_i \ketb{\psi_i}{\psi_i},\\
     & \mu_1 = \frac{p}{d^2} + 1 - p, \\
     & \mu_2 = \cdots = \mu_{d^2} = \frac{p}{d^2},\\
     & \mu_{d^2+1} = \cdots = \mu_{D^2} = 0.
\end{aligned}
\end{equation}
So the result in Eq.~(\ref{eq: depo_fidelity_function}) is
\begin{equation}
\begin{aligned}
    &F(p\MaxMixedState_{D\times D} + (1-p)\MaxEntangledState_{D}, p \frac{\1_{d\times d}}{d^2} \ot \MaxEntangledState_{D/d} + (1-p)\MaxEntangledState_{D}) \\
    =& \left[\tr \left(\sqrt{\sqrt{\sum_{i=1}^{D^2} \lambda_i \ketb{\psi_i}{\psi_i}} \sum_{i=1}^{D^2} \mu_i \ketb{\psi_i}{\psi_i} \sqrt{\sum_{i=1}^{D^2} \lambda_i \ketb{\psi_i}{\psi_i}} } \right)\right]^2 \\
    =& \left[ \tr \left( \sum_{i=1}^{D^2} \sqrt{\lambda_i \mu_i} \ketb{\psi_i}{\psi_i} \right) \right]^2 \\
    =& \left[ \sum_{i=1}^{D^2} \sqrt{\lambda_i \mu_i} \right]^2 \\
    =& \left[\sqrt{(\frac{p}{D^2} + 1 - p)(\frac{p}{d^2} + 1 - p)} + (d^2-1)\frac{p}{Dd}\right]^2.
\end{aligned}
\end{equation}
\end{proof}

\section{Sensitive analysis of \VQAname~for compressing information within quantum circuits}
\label{Sec:sensitive analysis}
\begin{figure}[!htbp]
    \centering
    \subfigure[Number Analysis]{
    \label{fig:circuit_number_impact}
    \includegraphics[width=0.3\textwidth]{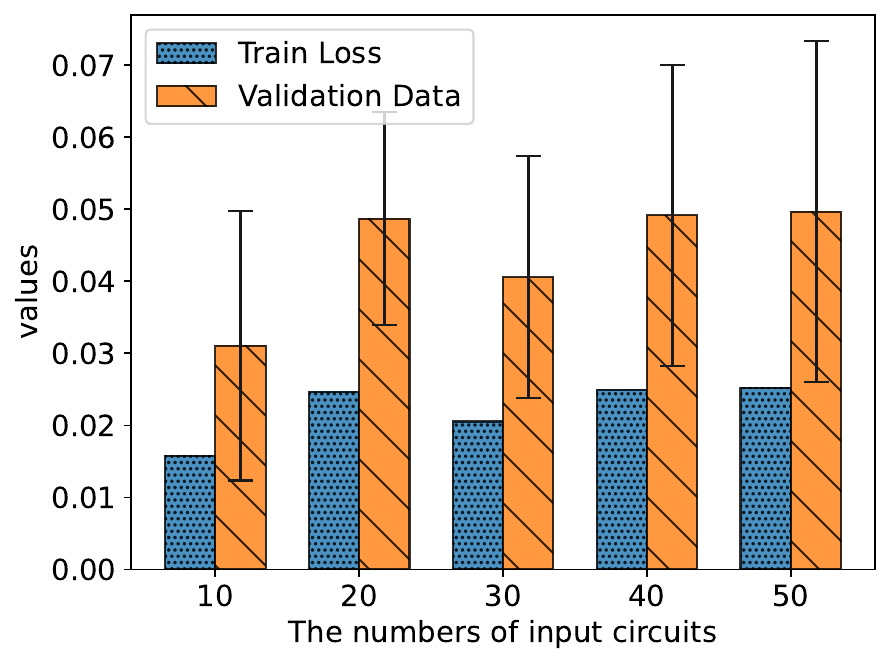}
    }
    \subfigure[Depth Analysis]{
    \label{fig:depth_impatct}
    \includegraphics[width=0.3\textwidth]{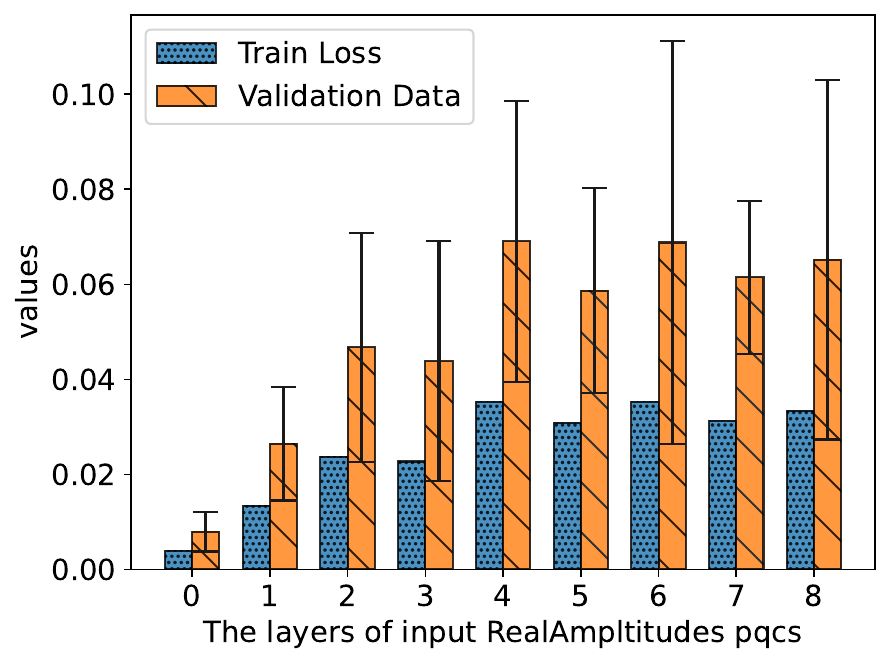}
    }
    \subfigure[Sigma Analysis]{
    \label{fig:sigma_analysis}
    \includegraphics[width=0.3\textwidth]{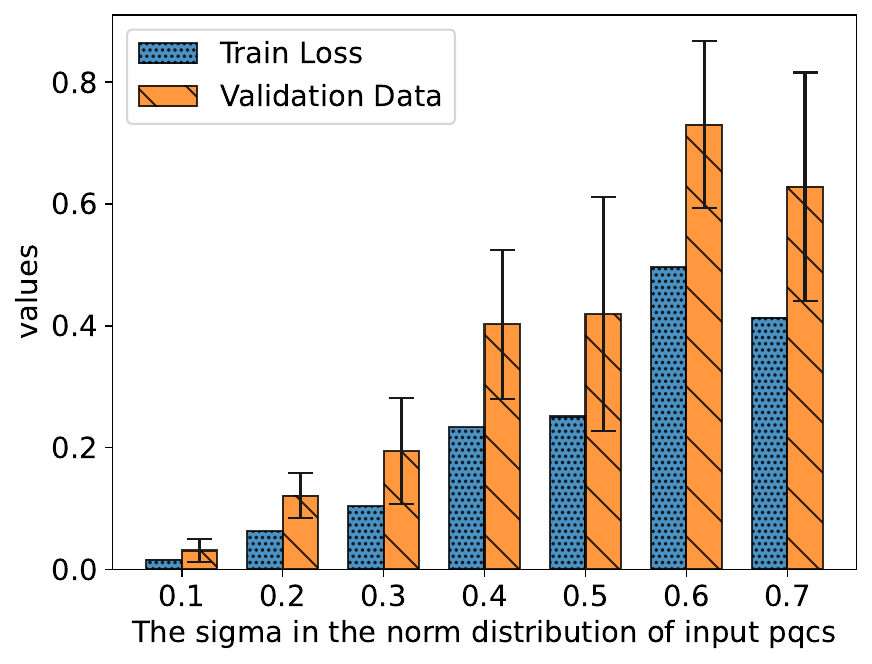}
    }
    \caption{Sensitive analysis on different parameters. We analyze the impacts of change in different input circuit datasets, including the number of circuits, the layers of each circuit, and the sigma in the normal distribution to generate control parameters in input circuits.}
    \label{fig:sentive-analysis}
\end{figure}

In this section, we experimentally analyze the impact of superparameters in \VQAname~for compressing information within quantum circuits. Fig.~\ref{fig:sentive-analysis} illustrates the performance obtained when changing some settings, including the number of input circuits, the parameters' distribution, and the number of qubits in circuits. 

The experiment shows that the loss function and validation values increase with the increase in the number of input circuits. The reason is that the mixed channel's rank increase as the input circuit increase. In Fig.~\ref{fig:circuit_number_impact}, the loss and validation are still under 0.07 when the number of input circuits increased to 50. In this experiment, we set the number of original and latent circuits as 4 and 3, use 1-layer ansatz to construct the input circuits, and the distribution to generate control parameters is $N(0,0.1)$.

In Fig.~\ref{fig:depth_impatct}, the results show that the loss function and validation values increase with the increase of the layers of ansatzes used in input circuits. The reason is also the increase of the rank of the mixed channel. In this experiment, we set the number of original and latent circuits as 4 and 3. and the distribution to generate control parameters is $N(0,0.1)$. In this experiment, we set the number of original and latent circuits as 4 and 3, use 20 circuits as the input, and the distribution to generate control parameters is $N(0,0.1)$.

Fig.~\ref{fig:sigma_analysis} change the values of $\sigma$ in the distribution $N(\mu, \sigma)$, we can find that as the $\sigma$ increases, the train and validation performance dramatically falls. This observation reveals that the rank of the input channel goes full as the number of circuits increases when the input circuits are all random unitaries. In this experiment, we set the number of original and latent circuits as 4 and 3, use 20 circuits as the input, use 1-layer ansatz to construct the input circuits, and the distribution to generate control parameters is $N(0,\sigma)$.

All the experiments show that the performance will be influenced dramatically as the rank of the input channel increases, which meets the description of the upper bound of \VQAname in Prop.~\ref{recovery_fidelity_bounded}.

\section{The diagram of designing swap circuit}
\label{append: swap_circuit}
The swap circuit construction issue, proposed in the Sec.~\ref{sec: validation}, is a crucial technique to obtain the channel $\Fc \ot \id$. For a $2n + 2m$ qubits system, swap the $[n+1, 2n]$ subsystem with the $[2n+1, 2n+m]$ subsystem. This problem can be transformed to an equivalent permutation problem: For a number list $L_1 := [a_1, \cdots, a_{2m}, b_1, \cdots, b_{2n}]$, find a permutation sequences $\{[x_i, y_j]|x_i,y_j \in L_1, x_i \neq y_j\}$ to get the number list $L_2 := [a_1, \cdots, a_m, b_1, \cdots, b_n, a_{m+1}, \cdots, a_{2m}, b_{n+1}, \cdots, b_{2n}]$.

For the swap circuit equivalent permutation problem, the following scheme:
\begin{itemize}
    \item {\em SWAP} $a_{2m}$ with $\{b_1,\cdots, b_n\}$;
    \item {\em SWAP} $a_{2m-1}$ with $\{b_1,\cdots, b_n\}$;
    \item $\cdots$;
\end{itemize}
The Eq.~(\ref{eq: swap_permutation}) illustrates the detail process of this scheme.
\begin{equation}
\label{eq: swap_permutation}
\begin{aligned}
     a_1, \cdots, a_{2m}&, b_1, \cdots, b_{2n}\\
    &\downarrow\\
    a_1, \cdots, a_{2m-1}, b_1, \cdots&, b_{n}, a_{2m}, b_{n+1}, \cdots, b_{2n}\\
    &\downarrow\\
    a_1, \cdots, a_{2m-2}, b_1, \cdots, &b_{n}, a_{2m-1}, a_{2m}, b_{n+1}, \cdots, b_{2n}\\
    &\downarrow\\
    &\cdots\\
    &\downarrow\\
    a_1, \cdots, a_{m}, b_1, \cdots, b_{n},& a_{m+1}, \cdots, a_{2m}, b_{n+1}, \cdots, b_{2n}.
\end{aligned}
\end{equation}

\bibliographystyle{apsrev4-2.bst}
\bibliography{sample.bib}

\begin{thebibliography}{32}%
\makeatletter
\providecommand \@ifxundefined [1]{%
 \@ifx{#1\undefined}
}%
\providecommand \@ifnum [1]{%
 \ifnum #1\expandafter \@firstoftwo
 \else \expandafter \@secondoftwo
 \fi
}%
\providecommand \@ifx [1]{%
 \ifx #1\expandafter \@firstoftwo
 \else \expandafter \@secondoftwo
 \fi
}%
\providecommand \natexlab [1]{#1}%
\providecommand \enquote  [1]{``#1''}%
\providecommand \bibnamefont  [1]{#1}%
\providecommand \bibfnamefont [1]{#1}%
\providecommand \citenamefont [1]{#1}%
\providecommand \href@noop [0]{\@secondoftwo}%
\providecommand \href [0]{\begingroup \@sanitize@url \@href}%
\providecommand \@href[1]{\@@startlink{#1}\@@href}%
\providecommand \@@href[1]{\endgroup#1\@@endlink}%
\providecommand \@sanitize@url [0]{\catcode `\\12\catcode `\$12\catcode `\&12\catcode `\#12\catcode `\^12\catcode `\_12\catcode `\%12\relax}%
\providecommand \@@startlink[1]{}%
\providecommand \@@endlink[0]{}%
\providecommand \url  [0]{\begingroup\@sanitize@url \@url }%
\providecommand \@url [1]{\endgroup\@href {#1}{\urlprefix }}%
\providecommand \urlprefix  [0]{URL }%
\providecommand \Eprint [0]{\href }%
\providecommand \doibase [0]{https://doi.org/}%
\providecommand \selectlanguage [0]{\@gobble}%
\providecommand \bibinfo  [0]{\@secondoftwo}%
\providecommand \bibfield  [0]{\@secondoftwo}%
\providecommand \translation [1]{[#1]}%
\providecommand \BibitemOpen [0]{}%
\providecommand \bibitemStop [0]{}%
\providecommand \bibitemNoStop [0]{.\EOS\space}%
\providecommand \EOS [0]{\spacefactor3000\relax}%
\providecommand \BibitemShut  [1]{\csname bibitem#1\endcsname}%
\let\auto@bib@innerbib\@empty
\bibitem [{\citenamefont {Liou}\ \emph {et~al.}(2008)\citenamefont {Liou}, \citenamefont {Huang},\ and\ \citenamefont {Yang}}]{liou2008modeling}%
  \BibitemOpen
  \bibfield  {author} {\bibinfo {author} {\bibfnamefont {C.-Y.}\ \bibnamefont {Liou}}, \bibinfo {author} {\bibfnamefont {J.-C.}\ \bibnamefont {Huang}},\ and\ \bibinfo {author} {\bibfnamefont {W.-C.}\ \bibnamefont {Yang}},\ }\href@noop {} {\bibfield  {journal} {\bibinfo  {journal} {Neurocomputing}\ }\textbf {\bibinfo {volume} {71}},\ \bibinfo {pages} {3150} (\bibinfo {year} {2008})}\BibitemShut {NoStop}%
\bibitem [{\citenamefont {Romero}\ \emph {et~al.}(2017)\citenamefont {Romero}, \citenamefont {Olson},\ and\ \citenamefont {Aspuru-Guzik}}]{romero2017quantum}%
  \BibitemOpen
  \bibfield  {author} {\bibinfo {author} {\bibfnamefont {J.}~\bibnamefont {Romero}}, \bibinfo {author} {\bibfnamefont {J.~P.}\ \bibnamefont {Olson}},\ and\ \bibinfo {author} {\bibfnamefont {A.}~\bibnamefont {Aspuru-Guzik}},\ }\href@noop {} {\bibfield  {journal} {\bibinfo  {journal} {Quantum Science and Technology}\ }\textbf {\bibinfo {volume} {2}},\ \bibinfo {pages} {045001} (\bibinfo {year} {2017})}\BibitemShut {NoStop}%
\bibitem [{\citenamefont {Wan}\ \emph {et~al.}(2017)\citenamefont {Wan}, \citenamefont {Dahlsten}, \citenamefont {Kristj{\'a}nsson}, \citenamefont {Gardner},\ and\ \citenamefont {Kim}}]{wan2017quantum}%
  \BibitemOpen
  \bibfield  {author} {\bibinfo {author} {\bibfnamefont {K.~H.}\ \bibnamefont {Wan}}, \bibinfo {author} {\bibfnamefont {O.}~\bibnamefont {Dahlsten}}, \bibinfo {author} {\bibfnamefont {H.}~\bibnamefont {Kristj{\'a}nsson}}, \bibinfo {author} {\bibfnamefont {R.}~\bibnamefont {Gardner}},\ and\ \bibinfo {author} {\bibfnamefont {M.}~\bibnamefont {Kim}},\ }\href@noop {} {\bibfield  {journal} {\bibinfo  {journal} {npj Quantum information}\ }\textbf {\bibinfo {volume} {3}},\ \bibinfo {pages} {1} (\bibinfo {year} {2017})}\BibitemShut {NoStop}%
\bibitem [{\citenamefont {Verdon}\ \emph {et~al.}(2018)\citenamefont {Verdon}, \citenamefont {Pye},\ and\ \citenamefont {Broughton}}]{verdon2018universal}%
  \BibitemOpen
  \bibfield  {author} {\bibinfo {author} {\bibfnamefont {G.}~\bibnamefont {Verdon}}, \bibinfo {author} {\bibfnamefont {J.}~\bibnamefont {Pye}},\ and\ \bibinfo {author} {\bibfnamefont {M.}~\bibnamefont {Broughton}},\ }\href@noop {} {\bibfield  {journal} {\bibinfo  {journal} {arXiv preprint arXiv:1806.09729}\ } (\bibinfo {year} {2018})}\BibitemShut {NoStop}%
\bibitem [{\citenamefont {Bondarenko}\ and\ \citenamefont {Feldmann}(2020)}]{bondarenko2020quantum}%
  \BibitemOpen
  \bibfield  {author} {\bibinfo {author} {\bibfnamefont {D.}~\bibnamefont {Bondarenko}}\ and\ \bibinfo {author} {\bibfnamefont {P.}~\bibnamefont {Feldmann}},\ }\href@noop {} {\bibfield  {journal} {\bibinfo  {journal} {Physical review letters}\ }\textbf {\bibinfo {volume} {124}},\ \bibinfo {pages} {130502} (\bibinfo {year} {2020})}\BibitemShut {NoStop}%
\bibitem [{\citenamefont {Huang}\ \emph {et~al.}(2020)\citenamefont {Huang}, \citenamefont {Ma}, \citenamefont {Yin}, \citenamefont {Tang}, \citenamefont {Dong}, \citenamefont {Chen}, \citenamefont {Xiang}, \citenamefont {Li},\ and\ \citenamefont {Guo}}]{huang2020realization}%
  \BibitemOpen
  \bibfield  {author} {\bibinfo {author} {\bibfnamefont {C.-J.}\ \bibnamefont {Huang}}, \bibinfo {author} {\bibfnamefont {H.}~\bibnamefont {Ma}}, \bibinfo {author} {\bibfnamefont {Q.}~\bibnamefont {Yin}}, \bibinfo {author} {\bibfnamefont {J.-F.}\ \bibnamefont {Tang}}, \bibinfo {author} {\bibfnamefont {D.}~\bibnamefont {Dong}}, \bibinfo {author} {\bibfnamefont {C.}~\bibnamefont {Chen}}, \bibinfo {author} {\bibfnamefont {G.-Y.}\ \bibnamefont {Xiang}}, \bibinfo {author} {\bibfnamefont {C.-F.}\ \bibnamefont {Li}},\ and\ \bibinfo {author} {\bibfnamefont {G.-C.}\ \bibnamefont {Guo}},\ }\href@noop {} {\bibfield  {journal} {\bibinfo  {journal} {Physical Review A}\ }\textbf {\bibinfo {volume} {102}},\ \bibinfo {pages} {032412} (\bibinfo {year} {2020})}\BibitemShut {NoStop}%
\bibitem [{\citenamefont {Du}\ and\ \citenamefont {Tao}(2021)}]{du2021exploring}%
  \BibitemOpen
  \bibfield  {author} {\bibinfo {author} {\bibfnamefont {Y.}~\bibnamefont {Du}}\ and\ \bibinfo {author} {\bibfnamefont {D.}~\bibnamefont {Tao}},\ }\href@noop {} {\bibinfo {title} {On exploring practical potentials of quantum auto-encoder with advantages}} (\bibinfo {year} {2021}),\ \Eprint {https://arxiv.org/abs/2106.15432} {arXiv:2106.15432 [quant-ph]} \BibitemShut {NoStop}%
\bibitem [{\citenamefont {Cerezo}\ \emph {et~al.}(2021{\natexlab{a}})\citenamefont {Cerezo}, \citenamefont {Sone}, \citenamefont {Volkoff}, \citenamefont {Cincio},\ and\ \citenamefont {Coles}}]{cerezo2021cost}%
  \BibitemOpen
  \bibfield  {author} {\bibinfo {author} {\bibfnamefont {M.}~\bibnamefont {Cerezo}}, \bibinfo {author} {\bibfnamefont {A.}~\bibnamefont {Sone}}, \bibinfo {author} {\bibfnamefont {T.}~\bibnamefont {Volkoff}}, \bibinfo {author} {\bibfnamefont {L.}~\bibnamefont {Cincio}},\ and\ \bibinfo {author} {\bibfnamefont {P.~J.}\ \bibnamefont {Coles}},\ }\href@noop {} {\bibfield  {journal} {\bibinfo  {journal} {Nature communications}\ }\textbf {\bibinfo {volume} {12}},\ \bibinfo {pages} {1} (\bibinfo {year} {2021}{\natexlab{a}})}\BibitemShut {NoStop}%
\bibitem [{\citenamefont {Cao}\ and\ \citenamefont {Wang}(2021)}]{cao2021noise}%
  \BibitemOpen
  \bibfield  {author} {\bibinfo {author} {\bibfnamefont {C.}~\bibnamefont {Cao}}\ and\ \bibinfo {author} {\bibfnamefont {X.}~\bibnamefont {Wang}},\ }\href@noop {} {\bibfield  {journal} {\bibinfo  {journal} {Physical Review Applied}\ }\textbf {\bibinfo {volume} {15}},\ \bibinfo {pages} {054012} (\bibinfo {year} {2021})}\BibitemShut {NoStop}%
\bibitem [{\citenamefont {Giovannetti}\ \emph {et~al.}(2008)\citenamefont {Giovannetti}, \citenamefont {Lloyd},\ and\ \citenamefont {Maccone}}]{giovannetti2008quantum}%
  \BibitemOpen
  \bibfield  {author} {\bibinfo {author} {\bibfnamefont {V.}~\bibnamefont {Giovannetti}}, \bibinfo {author} {\bibfnamefont {S.}~\bibnamefont {Lloyd}},\ and\ \bibinfo {author} {\bibfnamefont {L.}~\bibnamefont {Maccone}},\ }\href@noop {} {\bibfield  {journal} {\bibinfo  {journal} {Physical review letters}\ }\textbf {\bibinfo {volume} {100}},\ \bibinfo {pages} {160501} (\bibinfo {year} {2008})}\BibitemShut {NoStop}%
\bibitem [{\citenamefont {Bharti}\ \emph {et~al.}(2022)\citenamefont {Bharti}, \citenamefont {Cervera-Lierta}, \citenamefont {Kyaw}, \citenamefont {Haug}, \citenamefont {Alperin-Lea}, \citenamefont {Anand}, \citenamefont {Degroote}, \citenamefont {Heimonen}, \citenamefont {Kottmann}, \citenamefont {Menke} \emph {et~al.}}]{bharti2022noisy}%
  \BibitemOpen
  \bibfield  {author} {\bibinfo {author} {\bibfnamefont {K.}~\bibnamefont {Bharti}}, \bibinfo {author} {\bibfnamefont {A.}~\bibnamefont {Cervera-Lierta}}, \bibinfo {author} {\bibfnamefont {T.~H.}\ \bibnamefont {Kyaw}}, \bibinfo {author} {\bibfnamefont {T.}~\bibnamefont {Haug}}, \bibinfo {author} {\bibfnamefont {S.}~\bibnamefont {Alperin-Lea}}, \bibinfo {author} {\bibfnamefont {A.}~\bibnamefont {Anand}}, \bibinfo {author} {\bibfnamefont {M.}~\bibnamefont {Degroote}}, \bibinfo {author} {\bibfnamefont {H.}~\bibnamefont {Heimonen}}, \bibinfo {author} {\bibfnamefont {J.~S.}\ \bibnamefont {Kottmann}}, \bibinfo {author} {\bibfnamefont {T.}~\bibnamefont {Menke}}, \emph {et~al.},\ }\href@noop {} {\bibfield  {journal} {\bibinfo  {journal} {Reviews of Modern Physics}\ }\textbf {\bibinfo {volume} {94}},\ \bibinfo {pages} {015004} (\bibinfo {year} {2022})}\BibitemShut {NoStop}%
\bibitem [{\citenamefont {Grant}\ \emph {et~al.}(2018)\citenamefont {Grant}, \citenamefont {Benedetti}, \citenamefont {Cao}, \citenamefont {Hallam}, \citenamefont {Lockhart}, \citenamefont {Stojevic}, \citenamefont {Green},\ and\ \citenamefont {Severini}}]{grant2018hierarchical}%
  \BibitemOpen
  \bibfield  {author} {\bibinfo {author} {\bibfnamefont {E.}~\bibnamefont {Grant}}, \bibinfo {author} {\bibfnamefont {M.}~\bibnamefont {Benedetti}}, \bibinfo {author} {\bibfnamefont {S.}~\bibnamefont {Cao}}, \bibinfo {author} {\bibfnamefont {A.}~\bibnamefont {Hallam}}, \bibinfo {author} {\bibfnamefont {J.}~\bibnamefont {Lockhart}}, \bibinfo {author} {\bibfnamefont {V.}~\bibnamefont {Stojevic}}, \bibinfo {author} {\bibfnamefont {A.~G.}\ \bibnamefont {Green}},\ and\ \bibinfo {author} {\bibfnamefont {S.}~\bibnamefont {Severini}},\ }\href@noop {} {\bibfield  {journal} {\bibinfo  {journal} {npj Quantum Information}\ }\textbf {\bibinfo {volume} {4}},\ \bibinfo {pages} {65} (\bibinfo {year} {2018})}\BibitemShut {NoStop}%
\bibitem [{\citenamefont {Fisher}(1988)}]{misc_iris_53}%
  \BibitemOpen
  \bibfield  {author} {\bibinfo {author} {\bibfnamefont {R.~A.}\ \bibnamefont {Fisher}},\ }\href@noop {} {\bibinfo {title} {{Iris}}},\ \bibinfo {howpublished} {UCI Machine Learning Repository} (\bibinfo {year} {1988}),\ \bibinfo {note} {{DOI}: https://doi.org/10.24432/C56C76}\BibitemShut {NoStop}%
\bibitem [{\citenamefont {Chiribella}\ \emph {et~al.}(2015)\citenamefont {Chiribella}, \citenamefont {Yang},\ and\ \citenamefont {Huang}}]{chiribella2015universal}%
  \BibitemOpen
  \bibfield  {author} {\bibinfo {author} {\bibfnamefont {G.}~\bibnamefont {Chiribella}}, \bibinfo {author} {\bibfnamefont {Y.}~\bibnamefont {Yang}},\ and\ \bibinfo {author} {\bibfnamefont {C.}~\bibnamefont {Huang}},\ }\href@noop {} {\bibfield  {journal} {\bibinfo  {journal} {Physical review letters}\ }\textbf {\bibinfo {volume} {114}},\ \bibinfo {pages} {120504} (\bibinfo {year} {2015})}\BibitemShut {NoStop}%
\bibitem [{\citenamefont {Zhu}\ \emph {et~al.}(2023)\citenamefont {Zhu}, \citenamefont {Bai}, \citenamefont {Wang}, \citenamefont {Li},\ and\ \citenamefont {Chiribella}}]{zhu2023quantum}%
  \BibitemOpen
  \bibfield  {author} {\bibinfo {author} {\bibfnamefont {Y.}~\bibnamefont {Zhu}}, \bibinfo {author} {\bibfnamefont {G.}~\bibnamefont {Bai}}, \bibinfo {author} {\bibfnamefont {Y.}~\bibnamefont {Wang}}, \bibinfo {author} {\bibfnamefont {T.}~\bibnamefont {Li}},\ and\ \bibinfo {author} {\bibfnamefont {G.}~\bibnamefont {Chiribella}},\ }\href@noop {} {\bibfield  {journal} {\bibinfo  {journal} {Quantum Machine Intelligence}\ }\textbf {\bibinfo {volume} {5}},\ \bibinfo {pages} {27} (\bibinfo {year} {2023})}\BibitemShut {NoStop}%
\bibitem [{\citenamefont {Chiribella}\ \emph {et~al.}(2008)\citenamefont {Chiribella}, \citenamefont {D'Ariano},\ and\ \citenamefont {Perinotti}}]{chiribella2008transforming}%
  \BibitemOpen
  \bibfield  {author} {\bibinfo {author} {\bibfnamefont {G.}~\bibnamefont {Chiribella}}, \bibinfo {author} {\bibfnamefont {G.~M.}\ \bibnamefont {D'Ariano}},\ and\ \bibinfo {author} {\bibfnamefont {P.}~\bibnamefont {Perinotti}},\ }\href@noop {} {\bibfield  {journal} {\bibinfo  {journal} {Europhysics Letters}\ }\textbf {\bibinfo {volume} {83}},\ \bibinfo {pages} {30004} (\bibinfo {year} {2008})}\BibitemShut {NoStop}%
\bibitem [{\citenamefont {Cerezo}\ \emph {et~al.}(2021{\natexlab{b}})\citenamefont {Cerezo}, \citenamefont {Arrasmith}, \citenamefont {Babbush}, \citenamefont {Benjamin}, \citenamefont {Endo}, \citenamefont {Fujii}, \citenamefont {McClean}, \citenamefont {Mitarai}, \citenamefont {Yuan}, \citenamefont {Cincio} \emph {et~al.}}]{cerezo2021variational}%
  \BibitemOpen
  \bibfield  {author} {\bibinfo {author} {\bibfnamefont {M.}~\bibnamefont {Cerezo}}, \bibinfo {author} {\bibfnamefont {A.}~\bibnamefont {Arrasmith}}, \bibinfo {author} {\bibfnamefont {R.}~\bibnamefont {Babbush}}, \bibinfo {author} {\bibfnamefont {S.~C.}\ \bibnamefont {Benjamin}}, \bibinfo {author} {\bibfnamefont {S.}~\bibnamefont {Endo}}, \bibinfo {author} {\bibfnamefont {K.}~\bibnamefont {Fujii}}, \bibinfo {author} {\bibfnamefont {J.~R.}\ \bibnamefont {McClean}}, \bibinfo {author} {\bibfnamefont {K.}~\bibnamefont {Mitarai}}, \bibinfo {author} {\bibfnamefont {X.}~\bibnamefont {Yuan}}, \bibinfo {author} {\bibfnamefont {L.}~\bibnamefont {Cincio}}, \emph {et~al.},\ }\href@noop {} {\bibfield  {journal} {\bibinfo  {journal} {Nature Reviews Physics}\ ,\ \bibinfo {pages} {1}} (\bibinfo {year} {2021}{\natexlab{b}})}\BibitemShut {NoStop}%
\bibitem [{\citenamefont {Benedetti}\ \emph {et~al.}(2019)\citenamefont {Benedetti}, \citenamefont {Lloyd}, \citenamefont {Sack},\ and\ \citenamefont {Fiorentini}}]{benedetti2019parameterized}%
  \BibitemOpen
  \bibfield  {author} {\bibinfo {author} {\bibfnamefont {M.}~\bibnamefont {Benedetti}}, \bibinfo {author} {\bibfnamefont {E.}~\bibnamefont {Lloyd}}, \bibinfo {author} {\bibfnamefont {S.}~\bibnamefont {Sack}},\ and\ \bibinfo {author} {\bibfnamefont {M.}~\bibnamefont {Fiorentini}},\ }\href@noop {} {\bibfield  {journal} {\bibinfo  {journal} {Quantum Science and Technology}\ }\textbf {\bibinfo {volume} {4}},\ \bibinfo {pages} {043001} (\bibinfo {year} {2019})}\BibitemShut {NoStop}%
\bibitem [{\citenamefont {McClean}\ \emph {et~al.}(2018)\citenamefont {McClean}, \citenamefont {Boixo}, \citenamefont {Smelyanskiy}, \citenamefont {Babbush},\ and\ \citenamefont {Neven}}]{mcclean2018barren}%
  \BibitemOpen
  \bibfield  {author} {\bibinfo {author} {\bibfnamefont {J.~R.}\ \bibnamefont {McClean}}, \bibinfo {author} {\bibfnamefont {S.}~\bibnamefont {Boixo}}, \bibinfo {author} {\bibfnamefont {V.~N.}\ \bibnamefont {Smelyanskiy}}, \bibinfo {author} {\bibfnamefont {R.}~\bibnamefont {Babbush}},\ and\ \bibinfo {author} {\bibfnamefont {H.}~\bibnamefont {Neven}},\ }\href@noop {} {\bibfield  {journal} {\bibinfo  {journal} {Nature communications}\ }\textbf {\bibinfo {volume} {9}},\ \bibinfo {pages} {1} (\bibinfo {year} {2018})}\BibitemShut {NoStop}%
\bibitem [{\citenamefont {Kandala}\ \emph {et~al.}(2017)\citenamefont {Kandala}, \citenamefont {Mezzacapo}, \citenamefont {Temme}, \citenamefont {Takita}, \citenamefont {Brink}, \citenamefont {Chow},\ and\ \citenamefont {Gambetta}}]{kandala2017hardware}%
  \BibitemOpen
  \bibfield  {author} {\bibinfo {author} {\bibfnamefont {A.}~\bibnamefont {Kandala}}, \bibinfo {author} {\bibfnamefont {A.}~\bibnamefont {Mezzacapo}}, \bibinfo {author} {\bibfnamefont {K.}~\bibnamefont {Temme}}, \bibinfo {author} {\bibfnamefont {M.}~\bibnamefont {Takita}}, \bibinfo {author} {\bibfnamefont {M.}~\bibnamefont {Brink}}, \bibinfo {author} {\bibfnamefont {J.~M.}\ \bibnamefont {Chow}},\ and\ \bibinfo {author} {\bibfnamefont {J.~M.}\ \bibnamefont {Gambetta}},\ }\href@noop {} {\bibfield  {journal} {\bibinfo  {journal} {nature}\ }\textbf {\bibinfo {volume} {549}},\ \bibinfo {pages} {242} (\bibinfo {year} {2017})}\BibitemShut {NoStop}%
\bibitem [{\citenamefont {Wang}\ \emph {et~al.}(2016)\citenamefont {Wang}, \citenamefont {Yao},\ and\ \citenamefont {Zhao}}]{wang2016auto}%
  \BibitemOpen
  \bibfield  {author} {\bibinfo {author} {\bibfnamefont {Y.}~\bibnamefont {Wang}}, \bibinfo {author} {\bibfnamefont {H.}~\bibnamefont {Yao}},\ and\ \bibinfo {author} {\bibfnamefont {S.}~\bibnamefont {Zhao}},\ }\href@noop {} {\bibfield  {journal} {\bibinfo  {journal} {Neurocomputing}\ }\textbf {\bibinfo {volume} {184}},\ \bibinfo {pages} {232} (\bibinfo {year} {2016})}\BibitemShut {NoStop}%
\bibitem [{\citenamefont {Chalapathy}\ and\ \citenamefont {Chawla}(2019)}]{chalapathy2019deep}%
  \BibitemOpen
  \bibfield  {author} {\bibinfo {author} {\bibfnamefont {R.}~\bibnamefont {Chalapathy}}\ and\ \bibinfo {author} {\bibfnamefont {S.}~\bibnamefont {Chawla}},\ }\href@noop {} {\bibfield  {journal} {\bibinfo  {journal} {arXiv preprint arXiv:1901.03407}\ } (\bibinfo {year} {2019})}\BibitemShut {NoStop}%
\bibitem [{\citenamefont {Gondara}(2016)}]{gondara2016medical}%
  \BibitemOpen
  \bibfield  {author} {\bibinfo {author} {\bibfnamefont {L.}~\bibnamefont {Gondara}},\ }in\ \href@noop {} {\emph {\bibinfo {booktitle} {2016 IEEE 16th international conference on data mining workshops (ICDMW)}}}\ (\bibinfo {organization} {IEEE},\ \bibinfo {year} {2016})\ pp.\ \bibinfo {pages} {241--246}\BibitemShut {NoStop}%
\bibitem [{\citenamefont {ANIS}\ \emph {et~al.}(2021)\citenamefont {ANIS}, \citenamefont {Abby-Mitchell}, \citenamefont {Abraham},\ and\ \citenamefont {et. al.}}]{Qiskit}%
  \BibitemOpen
  \bibfield  {author} {\bibinfo {author} {\bibfnamefont {M.~S.}\ \bibnamefont {ANIS}}, \bibinfo {author} {\bibnamefont {Abby-Mitchell}}, \bibinfo {author} {\bibfnamefont {H.}~\bibnamefont {Abraham}},\ and\ \bibinfo {author} {\bibfnamefont {A.}~\bibnamefont {et. al.}},\ }\href {https://doi.org/10.5281/zenodo.2573505} {\bibinfo {title} {Qiskit: An open-source framework for quantum computing}} (\bibinfo {year} {2021})\BibitemShut {NoStop}%
\bibitem [{\citenamefont {Developer}(2021)}]{mq_2021}%
  \BibitemOpen
  \bibfield  {author} {\bibinfo {author} {\bibfnamefont {M.}~\bibnamefont {Developer}},\ }\href {https://gitee.com/mindspore/mindquantum} {\bibinfo {title} {Mindquantum, version 0.6.0}} (\bibinfo {year} {2021})\BibitemShut {NoStop}%
\bibitem [{\citenamefont {Wei}\ \emph {et~al.}(2018)\citenamefont {Wei}, \citenamefont {Xin},\ and\ \citenamefont {Long}}]{wei2018efficient}%
  \BibitemOpen
  \bibfield  {author} {\bibinfo {author} {\bibfnamefont {S.-J.}\ \bibnamefont {Wei}}, \bibinfo {author} {\bibfnamefont {T.}~\bibnamefont {Xin}},\ and\ \bibinfo {author} {\bibfnamefont {G.-L.}\ \bibnamefont {Long}},\ }\href@noop {} {\bibfield  {journal} {\bibinfo  {journal} {Science China Physics, Mechanics \& Astronomy}\ }\textbf {\bibinfo {volume} {61}},\ \bibinfo {pages} {1} (\bibinfo {year} {2018})}\BibitemShut {NoStop}%
\bibitem [{\citenamefont {Jamio{\l}kowski}(1972)}]{jamiolkowski1972linear}%
  \BibitemOpen
  \bibfield  {author} {\bibinfo {author} {\bibfnamefont {A.}~\bibnamefont {Jamio{\l}kowski}},\ }\href@noop {} {\bibfield  {journal} {\bibinfo  {journal} {Reports on Mathematical Physics}\ }\textbf {\bibinfo {volume} {3}},\ \bibinfo {pages} {275} (\bibinfo {year} {1972})}\BibitemShut {NoStop}%
\bibitem [{\citenamefont {Choi}(1975)}]{choi1975completely}%
  \BibitemOpen
  \bibfield  {author} {\bibinfo {author} {\bibfnamefont {M.-D.}\ \bibnamefont {Choi}},\ }\href@noop {} {\bibfield  {journal} {\bibinfo  {journal} {Linear Algebra and its Applications}\ }\textbf {\bibinfo {volume} {10}},\ \bibinfo {pages} {285} (\bibinfo {year} {1975})}\BibitemShut {NoStop}%
\bibitem [{\citenamefont {Cerezo}\ and\ \citenamefont {Coles}(2020)}]{cerezo2020impact}%
  \BibitemOpen
  \bibfield  {author} {\bibinfo {author} {\bibfnamefont {M.}~\bibnamefont {Cerezo}}\ and\ \bibinfo {author} {\bibfnamefont {P.~J.}\ \bibnamefont {Coles}},\ }\href@noop {} {\bibfield  {journal} {\bibinfo  {journal} {arXiv e-prints}\ ,\ \bibinfo {pages} {arXiv}} (\bibinfo {year} {2020})}\BibitemShut {NoStop}%
\bibitem [{wjq()}]{wjqcae}%
  \BibitemOpen
  \href@noop {} {}\bibinfo {howpublished} {\url{https://github.com/linke-quantum/QCAE-master}}\BibitemShut {NoStop}%
\bibitem [{\citenamefont {An}\ and\ \citenamefont {Cho}(2015)}]{an2015variational}%
  \BibitemOpen
  \bibfield  {author} {\bibinfo {author} {\bibfnamefont {J.}~\bibnamefont {An}}\ and\ \bibinfo {author} {\bibfnamefont {S.}~\bibnamefont {Cho}},\ }\href@noop {} {\bibfield  {journal} {\bibinfo  {journal} {Special lecture on IE}\ }\textbf {\bibinfo {volume} {2}},\ \bibinfo {pages} {1} (\bibinfo {year} {2015})}\BibitemShut {NoStop}%
\bibitem [{\citenamefont {Watrous}(2018)}]{watrous2018theory}%
  \BibitemOpen
  \bibfield  {author} {\bibinfo {author} {\bibfnamefont {J.}~\bibnamefont {Watrous}},\ }\href@noop {} {\emph {\bibinfo {title} {The theory of quantum information}}}\ (\bibinfo  {publisher} {Cambridge university press},\ \bibinfo {year} {2018})\BibitemShut {NoStop}%
\end{thebibliography}%
\end{document}